\documentclass{JHEP3}
\usepackage{amsmath}
\usepackage{amsfonts}
\usepackage{amssymb}
\usepackage{epsfig}
\def\beq{\begin{equation}}
\def\beql#1{\begin{equation}\label{eq:#1}}
\def\eeq{\end{equation}}

\newcommand{\tr}{{\rm tr}~ }
\newcommand{\comment}[1]{}

\newcommand{\pasl}{\pa\kern-.55em /}

\newcommand{\ksl}{k\kern-.55em /}

\DeclareFixedFont{\xiiss}{OT1}{cmss}{m}{n}{12}
\DeclareFixedFont{\ixss}{OT1}{cmss}{m}{n}{9}
\DeclareFixedFont{\cmrnine}{OT1}{cmr}{m}{n}{9}
\newcommand{\field}[1]{\mathbb{#1}}
\newcommand{\BC}{{\field C}}

\newcommand{\BZ}{{\field Z}}

\newcommand{\CCs}{\hbox{\ixss C\kern-.4emI}}
\newcommand{\ZZs}{\hbox{\ixss Z\kern-.4emZ}}

\newcommand{\diag}{\hbox{diag}}

\newcommand{\littlefig}[2]{
 \epsfxsize=#2in
 \epsfbox{#1}
}

\title{Solving matrix models using holomorphy}

\author{David Berenstein \thanks{dberens@ias.edu}\\
 School of Natural Sciences,
Institute for Advanced Study, Princeton, NJ 08540, USA }

\abstract{We investigate the relationship between 
supersymmetric gauge theories with moduli 
spaces and matrix models. Particular attention is given to 
situations where the moduli space gets quantum corrected. These 
corrections are
controlled by holomorphy. It is argued that these quantum  
deformations give rise to
non-trivial relations for generalized resolvents 
that must hold in the associated matrix model.
 These relations allow to  
solve a sector of the associated matrix model in a similar way to
a one-matrix model,  by studying a curve that encodes the 
generalized resolvents. At the level of loop equations for the matrix
model, the situations with a moduli space can sometimes be considered as a
degeneration of an infinite set of linear equations, 
and the quantum moduli 
space encodes the consistency conditions for these equations
to have a solution.}


\begin{document}

\section{Introduction}

Recent developments on supersymmetric gauge theories have shown
that there is a deep relationship between SUSY theories that admit a large $N$
limit and
zero dimensional matrix models \cite{DV,DV2,DV3}.
The
connection between these seemingly different physical systems arose from
the study of topological string theory amplitudes for open
strings, but now there are purely field-theoretic arguments that show
this connection.
Of the two available proofs \cite{DGLVZ,CDSW} of the correspondence
(for a $U(N)$ gauge theory),
 the most compelling one was given in the
paper \cite{CDSW}, as it gave a full dictionary between loop operators
on the matrix model and certain elements of the chiral ring of the
gauge theory, as well as
a proof that can be argued to be valid non-perturbatively.
Part of the setup  includes a correspondence between the
 gaugino condensate of the SUSY gauge theory and the
't Hooft coupling of the matrix model \cite{DV3}.

Having this connection makes it possible to solve for the vacuum
structure of the supersymmetric gauge field theory if one knows
how to solve the matrix model in the large $N$ limit.
We will refer the reader to \cite{DGZ,GM} for reviews of matrix 
models. A more recent list of solvable models appears in \cite{Ka}

The most successful and studied example to date has been given by
deforming the $U(N)$ ${\cal N}=2$ gauge theory, whose solution was given by 
Seiberg and Witten \cite{SW} to an ${\cal N}=1$
gauge theory by adding a superpotential $\tr(V(X))$, where $X$ is
the chiral field superpartner of the vector multiplet and $V(X)$
is a polynomial in $X$, see for example \cite{CV, DV3, 
Chek, F, F2, Fuji, Gop, IM, CDSW, CSW}. This theory is
related to the one matrix model with potential $V(X)$. The
classical vacua of the gauge theory are described by distributing
eigenvalues on the roots of the polynomial $V'(X)$ giving rise to
a theory whose classically unbroken gauge group is $\prod U(N_i)$,
where
 there are $N_i$ eigenvalues in the $i$-th root of the polynomial
 $V'(X)$.

Similarly, in the matrix model, the vacua are also described by
distributing (an infinite number of) eigenvalues over the
classical saddle points of the matrix model. Each of these
classical points in the eigenvalue plane becomes a cut in the
eigenvalue plane due to quantum effects, when one considers the
spectral curve of the matrix model. For each cut there is an
associated 't Hooft coupling which counts how many eigenvalues end
up in the given cut. This information is related to a partial
gaugino condensate \cite{DV3}, whose holomorphic definition was
given nonperturbatively in \cite{CDSW}.

One of the advantages of this new method of computing gauge theory
results is that it does not assume S-duality like symmetries to
solve the theory, so it can provide tests of these highly
non-trivial symmetries. For example, aspects of the ${\cal N}=1^*$
vacuum structure have been studied in this way \cite{DHKS,DHKS2}, but a general
solution of the vacua of the theory in terms of matrix models is
not yet available.
 \footnote{This theory can also be solved by
first turning on a mass term deformation that gives rise to an
${\cal N}=2$ gauge theory, and finding the associated integrable
system \cite{DW,DHK,Hollow}. In principle, it is possible to evaluate all of
the holomorphic information by going to the appropriate point in
the moduli space of vacua.}

 Part of the failure to obtain the full solution of the matrix model
comes
from the fact that the classical ${\cal N}=1^*$ gauge theory vacua
is described by a decomposition into irreducible representations
of $SU(2)$, and there are an infinite number of such
representations. In the associated matrix model, this translates
into an infinite number of classical saddle points for the
eigenvalues of one of the matrices, let us call it $X$, and to
each such classical saddle point for the eigenvalue $x$ of $X$ one
can associate a cut in a spectral curve of the quantum theory (at
least, this is what experience dictates from the one-matrix
model). In particular, one has a potentially infinite number of
cuts. This suggests that one can not solve the loop equations of
the matrix model in a straightforward fashion, as there will be an
infinite number of unknown parameters: one can take these
parameters to be the number of eigenvalues in each cut $N_i$, but
more to the point, the one point functions $<X^n>$ depend
classically on all of the $N_i$. Thus, there are no recursion
relations which solve for the one point functions of the loop
operators $<X^n>$.

The technical issue is how to effectively truncate the problem to
finitely many cuts. In the one matrix model this is automatic, but
in the ${\cal N}=1^*$ theory one has to do this by hand. A related
problem in complex analysis can be phrased as follows: given a
Laurent series of an analytic function $f(z)=\sum a_i z^{-i}$ at
infinity, what are the constraints on the variables $a_i$ that
guarantee that $f(z)$ has a finite number of cuts in the interior
of the complex plane? The constraints involve an infinite number
of the coefficients of $f$ at the same time, since multiplying
$f(z)$ by any polynomial function of $z$ will always give a
function with the same properties. Looking at it from  this point
of view it also looks like the problem has no solution.

The current literature solves the abstract problem by making a clever
anzatz for the one-cut solution (or multi-cut solutions
 where all of the cuts are
correlated, see for example \cite{KKN, DHKS,DHKS2,DHK2,Man}),
that leads to studying functions on an elliptic curve, but it does
not solve in any sense a {\em general} multi-cut solution.

In this paper we will describe a new technique to solve (only
partially) similar matrix models based on getting information from
the gauge theory first. Since the gauge theory and the matrix
model data are in some sense equivalent, this statement might seem
paradoxical. We will concern ourselves with situations where the
superpotential is non-generic and there is the possibility of
having a moduli space of vacua. In particular, the allowed
deformations of the moduli space give a finite number of
parameters which determine part of the vacuum structure. In the
matrix model these turn out to be situations where the system of
loop equations is degenerate in some sense, a statement which will
be explained in detail later in  the paper, and there are more
free parameters than at a generic case. The main question we want
to address is: what does the moduli space tell us about the matrix
model?

To avoid undue suspense, we find that under some circumstances the
existence of a moduli space of  vacua produces an {\em integrable
sector} of the matrix model, this is, part of the problem behaves
like a one matrix model: one can find an infinite number of one
point functions with a finite amount of data. The finite amount of
data is exactly the one that characterizes the possible
deformations of the moduli space that are allowed by holomorphy.

To show the new technique, we will study two examples. First, a
toy example which reduces to the one matrix model, which we
understand very well. It will just be an illustration of the
matrix model techniques used.
 Then we will study a
three matrix model closely related to the ${\cal N}=1^*$ gauge
theory which has special properties that insure that certain
objects in the matrix model only have a finite number of cuts
automatically. This happy coincidence is tied to the fact that
these particularly gauge theories can be geometrically engineered.
We will in this paper always consider a gauge field theory which
can be geometrically engineered by placing branes on a singular
Calabi-Yau (CY) geometry. From this point of view we are exploring
 questions that relate the CY geometry to a matrix model.

The basic idea is that if we place fractional branes at the
singularities of the geometry, then the CY geometry will be
deformed by geometric transitions so that the singularities with
fractional branes on them are 'resolved'. This will be exactly like in the
Klebanov-Strassler geometry \cite{KS,GV}, repeated many times at many 
singularities, as in
\cite{CIV,CFIKV}.

If we insert a probe brane in the bulk (in the presence of the
fractional branes) then it's moduli space should correspond to the
deformed CY geometry. This idea is very common in the literature, see for 
example \cite{KS,GT,GT2,Bq}.
Fortunately, the shape of the CY can be
studied directly from the field theory \cite{BL,Brev}
 so it is possible to start
with a classical gauge field theory and produce certain CY
geometries, as opposed to start with the geometry and guess the field 
theory. If we place a brane in the bulk and compute it's
moduli space, the moduli space of the brane will be the CY
geometry. Of course, not every superpotential will lead to a CY
geometry, but for those that do one can make progress in
understanding the gauge theory questions by studying the geometry
(moduli space of the theory).
Also, one should be able to understand aspects of the matrix model by 
studying the geometry.

The main advantage with this setup is that the problem is controlled 
by holomorphy:
the possible deformations of the CY geometry that result from
placing fractional branes at singularities are not arbitrary, so
there are only a finite number of unknowns. This is exactly what
we are looking for, a situation where the potentially infinite
number of variables (cuts in the matrix model)
 is reduced to finitely many (we are assuming that
there is one cut per quantum deformation parameter, a statement which
will be justified later).

We want to study exactly what matrix model information we can get
from the gauge theory moduli space. We will use a proposal first
presented in \cite{Bq} to calculate the quantum moduli space from
the matrix model. The relation between the moduli space and how to
obtain it from the matrix model still lacks a proof, so it should
be treated as a conjecture that needs justification. However, one
can make predictions based on the calculation of the deformed
moduli space from the matrix model, and the one expected from
holomorphy. Some of these predictions can be verified directly by
manipulating the loop equations of the matrix model, and we will
use this calculation as a consistency check of the proposal.

In our examples, we will not solve the
matrix model completely (find all one point functions from a
finite amount of data). We will use the word `solve' in the
following more restricted sense: it is possible to determine an
infinite number of one point functions of the matrix model with a
finite amount of data. In essence, we obtain similar results to a one 
matrix model.

The main example we will study 
is a gauge theory that will have three matrices
$X,Y,Z$ in the adjoint of $U(N)$, and a superpotential of the
general form $$ \tr(XYZ-q XZY+ V(X)) $$ with the restriction
$q^n=1$, and a restriction on the form of $V(X)$ which depends on
$n$. Similar systems have been studied  
in \cite{D,DF,BJL,DHK2,Hollow, Man} and are interesting on their own right as
they encode structure of interesting 
deformations of ${\cal N}=4$ gauge theories \cite{LS}.
The  restrictions placed on the superpotential guarantee
 that there is a
CY geometry associated to the gauge theory \cite{D,DF,BJL}, and similar 
examples studied in \cite{CIV,CKV}.
 Also,  after a linear
change of variables on $X$, we can get for $q\neq 1$ an equivalent
superpotential of the form
$$
\tr(XYZ-q XZY+ \tilde V(X)+ m^2 YZ)
$$
The ${\cal N}=1^*$ gauge theory has the same matter content and
the potential
$$
\tr(XYZ- XZY+\tilde m^2 X^2 + m^2 YZ)
$$
so by varying $q$ we can in principle get arbitrarily close to
the ${\cal N}=1^*$ gauge theory. This particular
model has been studied extensively, also 
in relation to AdS/CFT \cite{PS,ADK}.
This is the main physical motivation to
write this paper: to approach the problem of solving the
${\cal N}=1^*$ vacuum structure from the matrix models,
 with an arbitrary number of 
cuts, as well as the related $q$ deformed systems. 
It is important to notice that setting $q\neq 1$ means that the conformal 
field theory with $V=0$ has only $N=1$ SUSY in four dimensions.
Therefore, techniques that depend on the quantum corrections 
being calculable 
in ${\cal N}=2$ theories, which then are softly broken 
to ${\cal N}=1$ can break down 
because there is less supersymmetry protecting the system. 
In particular, to use integrable systems like in ${\cal N}=2$ gauge theories,
but with less supersymmetry \cite{Hollow}, 
one actually needs a proof that the moduli space does not get lifted by 
quantum corrections.
These matrix models are also interesting on their own right, 
see for example \cite{Kos2}, so we can also gain insight into the problem 
of solving matrix models by using four dimensional SUSY gauge theory 
information.

\section{The one matrix model}\label{sec:1matrix}

In this short section we will show how the analysis of a quantum
moduli space can solve the one matrix model.  The whole purpose of
this section is to show that a quantum deformed CY geometry
contains information about loop equations of the matrix model.

Consider the  CY threefold  geometry
\begin{equation}\label{eq:CY1}
uv= (w- P(z))(w+P(z))
\end{equation}
where $P(z)$ is a polynomial of degree $n$ ($P(z)=\sum_{i=0}^n a_i
z^i$), and which results from deforming the $A_1$ singularity
geometry $uv = w^2$. This procedure removes the codimension two
singularities and leaves only codimension three singularities
behind, which for generic $P(z)$ are conifold singularities.

The gauge theory associated to D-branes on this geometry has been 
studied in \cite{CIV,CKV,CFIKV,CV}, and it is given by a quiver
diagram with two nodes, representing the affine dynkin diagram 
$\hat A_1$.

\littlefig{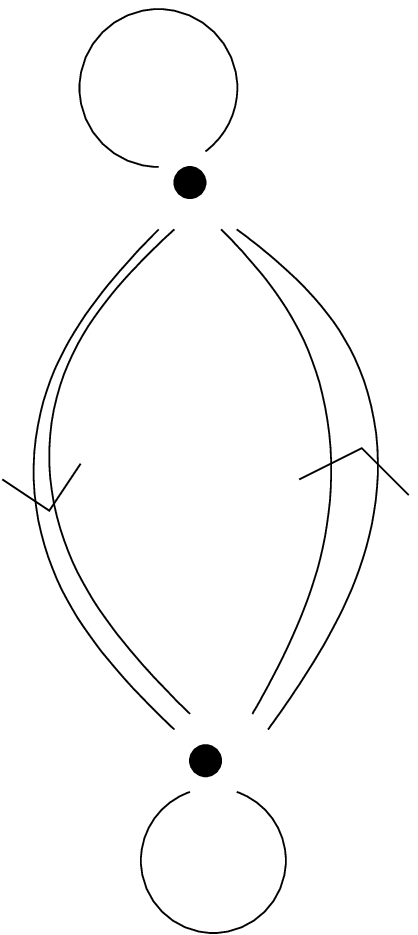}{1}

Because there are two nodes in the quiver theory, one has general
gauge group $U(N)\times U(M)$. The superpotential of the theory is
given by
\begin{equation}\label{eq:super1}
W = \tr( Z A_1B_1 - B_1 A_1 \tilde Z - ZA_2B_2+B_2A_2 \tilde Z)
+\tr(V(Z) - V(\tilde Z)
\end{equation}
and in the case $V(x)=0$ the superpotential is the one that results
from an ${\cal N}=2$ supersymmetric gauge theory. $P(x)$ as a
polynomial is determined by the following equation $2P(x)=V'(x)$,
so one can relate the shape of the geometry and the gauge theory
parameters in the superpotential.

A brane in the bulk will have a theory characterized by the gauge
group $U(1)\times U(1)$. This is the correct fractional brane
content for a bulk brane in the orbifold geometry, and these
numbers do not change when we deform the geometry.

Also, if we take $M=0$, we see that we obtain a gauge theory with the same
matter content as the ${\cal N}=2$, $U(N)$ gauge theory, which has been
deformed by the potential $V(Z)$. Thus this particular theory is a
sub-theory of the one defined by the geometry.

This $U(N)$ does not have a moduli space of vacua, instead it has
a discrete set of vacua. These
 vacua at the classical level
are built by distributing the $N$
eigenvalues of $Z$ into the roots of $P(Z)$.
If there are $N_i$
eigenvalues at the $r_i$ root, the low energy effective (classical) theory
is given by pure gauge theory, with gauge group $\prod_i U(N_i)$.
Quantum effects lead to confinement of the $SU(N_i)$ groups, so in the
infrared of the quantum theory
we end up with a gauge group $\prod(U(1))$ over the roots
which have at least one eigenvalue.

To solve for the structure of the vacuum, we want to find the vevs
of all of the elements of the chiral ring.
 The chiral ring is generated by the following combinations
\begin{eqnarray}
\frac 1 {32\pi^2}tr(Z^nW_\alpha W^\alpha)\\
tr(Z^nW_\alpha)\\
tr(Z^n)
\end{eqnarray}
The ordering of $W$ inside the trace does not matter, because any
other ordering can be obtained by the addition of terms of the
form $\bar D M$, so at the level of the chiral ring they give the
same cohomology class. For our purposes, the elements of the
chiral ring which have a simple interpretation in a matrix model
are the ones given by $tr(Z^nW_\alpha W^\alpha)$. Using the
generalized Konishi anomaly for the variations $\delta Z \sim Z^kW_\alpha
W^\alpha$, Cachazo et al. \cite{CDSW} were able to show that these satisfy the
same equations as the loop equations for the one matrix model with
potential $V(X)$. This is
\begin{equation}
tr(V'(Z) Z^k W_\alpha W^\alpha) =\frac{1}{32\pi^2} \sum(tr(Z^{k-1-i}W_\alpha
W^\alpha)tr(Z^iW_\alpha W^\alpha)
\end{equation}
If we call $O_i = \frac1{32\pi^2}tr(Z^iW_\alpha W^\alpha)$, then the
equations
above read
\begin{equation}
tr(\sum a_j Z^{k+j} W_\alpha W^\alpha) =\frac{1}{32\pi^2}
\sum(tr(Z^{k-1-i}W_\alpha
W^\alpha)tr(Z^iW_\alpha W^\alpha)
\end{equation}
or equivalently
\begin{equation}
\sum a_j O_{k+j} = \sum_i O_{k-i-1} O_i
\end{equation}
This equation can be read as  a recursion relation for the vevs
$O_{n+k}$ for $k\geq 0$ in terms of the vevs $O_{0},\dots,
O_{n-1}$. All of these equations can be put in one single equation
in terms of a generating function for the $O_i$, defined as
follows
\begin{equation}
R(u) = \sum O_k u^{-k-1}
\end{equation}
so that
\begin{equation}{\label{eq:loop1matrix}}
V'(u)R(u) = R(u)^2 +f(u)
\end{equation}
where $f$ is an unknown polynomial of degree $n-1$. This is
necessary to seed the initial conditions for the recursion
relation. Given $f$ it straightforward to solve for $R(u)$. $f$ in
the quantum theory is determined by how we choose to distribute
the eigenvalues of $Z$ on the different roots of the classical
potential, and it encodes the different possible vacua of the
theory.

These equations above are the same equations that can be derived
from a one matrix model with potential $V(X)$.
 This is, we have the
matrix model
\begin{equation}
\int [dX] \exp (-N\mu^{-1} \tr(V(X)))
\end{equation}
which is to be solved in the planar limit in the large $N$ limit.
We can write the loop operators
\begin{equation}
<X^k> = \frac 1N\tr(X^k)
\end{equation}
and the loop equations for the one matrix model
\begin{equation}
<V'(X)X^k> = \mu \sum_i<X^{k-1-i}><X^i>
\end{equation}
So it is immediate to see
 that they have structurally the same form as the ones given
above (by setting $O_i\sim \mu<X^i>$).
 In the matrix model
we have the normalization condition $<X^0> =<1> =1$, which
identifies $\mu\sim \frac1{32\pi^2}<W_\alpha W^\alpha>=S$,
the gaugino condensate.
It is the identification between loop equations of a matrix model
and the chiral ring constraints that allow us to relate these two
very different mathematical problems.

Now, let us use the quantum moduli space technique to rederive the same loop
equations from a different point of view. The idea is to study the
gauge theory with potential \ref{eq:super1}, with one brane in the
bulk, this is, with gauge group $U(N+1)\times U(1)$.

The classical moduli space is obtained by solving the F-term
constraints. The constraints for the $A,B$ give us
\begin{equation}
zA_i - A_i\tilde z =0\quad B_i z-\tilde z B_i=0
\end{equation}
and since $\tilde z$ is a scalar, for $A,B$ to get a non-zero vev
and give us a brane in the bulk, we need that one of the
eigenvalues of $z$ is equal to $\tilde z$, and that $A,B$ be
eigenvectors of the matrix $z$ with eigenvalue $\tilde z$.

We can diagonalize $z$, and single out the ($1\times 1$) block
matrices corresponding to the eigenvalue $\tilde z$. With this
convention
\begin{equation}
z= \diag(\tilde z, z_1 ,\dots z_n),\quad B= (b_i,0,\dots,0)\quad
A_i=\begin{pmatrix}a_i\\ 0\\\vdots\\0\end{pmatrix}
\end{equation}

The other eigenvalues will be constrained by the equations
\begin{equation}
A_1B_1-A_2B_2 = V'(z)\quad B_1A_1-B_2A_2=V'(\tilde z)
\end{equation}
which result in $V'(z_i)=0$ and
\begin{equation}
b_1a_1-b_2a_2 = V'(\tilde z)
\end{equation}
From here, the classical eigenvalues $z_i$ which were not singled out
will be distributed
only along the saddle points of the potential $V$.

Now, to find the moduli space we need to write the constraints
between the gauge invariant variables (with respect to the group
$U(N+1)$) given by $\tilde z$ and the combinations
\begin{equation}
u=a_1b_2, v=a_2b_1, w = \frac{a_1b_1+a_2b_2}{2},
\Omega=\frac{a_1b_1-a_2b_2}{2}
\end{equation}
 Clearly, the F-terms imply $\Omega=V'(\tilde z)/2$.
 And we can also use the algebraic relation
\begin{equation}
w^2 = \Omega^2+uv = \frac{V'(\tilde z)^2}{4}+uv
\end{equation}
to obtain the equations describing the moduli space given by the
geometry \ref{eq:CY1}.

The classical moduli space is given by the single equation in four
variables $w, u, v, \tilde z$
\begin{equation}
w^2-uv -\frac{V'(\tilde z)^2}{4}=0
\end{equation}
and it has branes located at the singularities of the geometry.
Because there are singularities, it is possible to deform the moduli
space and remove the singularities. These deformations are
controlled by holomorphy.

We should expect that the quantum deformed moduli space is of the
form
\begin{equation}
w^2-uv -\frac{V'(\tilde z)^2}{4} = \hbox{Quantum deformations}=Q
\end{equation}
where the right hand side is a polynomial involving only positive
powers of the confining scale $\Lambda_{SU(N+1)}$, and which is
polynomial in the coefficients of $P$. This is the expected result from
holomorphy \cite{Snal,Sexact,Shol}.

The particular theory at hand
has an $SU(2)$ symmetry under which $A_1,A_2$ and
$B_1,B_2$ are doublets, while $z,\tilde z$ are singlets. $w,u,v$
together form a triplet, and $\Omega$ is a singlet. From these
symmetries one can see that the quantum deformations will be
independent of $w,u,v$, since terms of degree one in these
variables that could appear in the polynomial $Q$ above are not in
a singlet representation of $SU(2)$. Hence $Q$ can only be given
by a polynomial in $z$. The deformations should also be such that they
are subleading with respect to the deformations induced by
changing the couplings in the superpotential. If $V'(z)$ is of
degree $n$, and we change the functional form of $V'$ we can vary
the coefficients of $z^n, z^{n+1}, \dots, z^{2n}$ independently of
each other. The deformation should thus be of the form $f(z)$, with $f$
of degree smaller than $n$ \cite{CIV}.

It turns out that the number of parameters in the deformation $f$
is equal to the number of roots of $V'$. For each of these roots
there is a conifold singularity, and each of these singularities
can be deformed away by placing fractional branes at the given
conifold, thus inducing a geometric transition which is of the same type as 
the one in \cite{KS}. We get in the end
the same number of deformation parameters
as there are singularities in the
geometry.

Now we want to see that these deformations can be related to the
loop equations of the matrix model with potential $V$.

For this, we need to know how to derive the quantum moduli space of
the gauge field theory from the matrix model point of view. This has
been previously argued for in \cite{Bq}.

We need to consider a multi-matrix model with the same field content as
the gauge theory, and we take ${\tilde N}\to \infty$. However,
 we fix the matrix model so that the number of moduli stays fixed.
This is, we will consider a matrix model with one probe brane in the
bulk singled out.
We do this with a matrix
model potential
given by the classical superpotential of the theory.
$$
\int ([dz][d\tilde z][d A][d B])' \exp(W)
$$
where the prime indicates that we leave some matrix fields unintegrated
(exactly those that are massless perturbatively in the gauge theory).
Since in the matrix model we do a large $N$ expansion, we are going
to get to the results by analyzing the saddle point equations for
the variables $\tilde z$,
and what we called $a_i,b_i$ perturbatively above. Also, one of the
eigenvalues of $z$ in the gauge theory is singled out to be equal to
$\tilde z$ perturbatively. This eigenvalue will be called $z_0$ in the
matrix model. At least naively, we can ignore the effect of the matrix 
probe on the large $N$ matrix condensate because it would give rise 
to a $1/N$ effect.

On going to a basis of eigenvalues for $z$, and including the
Vandermonde determinant  we find the following
integral to perform in the large $\tilde {N}$
limit:
\begin{eqnarray}
&\int \prod_{i\neq 0}[d\lambda_i][dA_{12}^i][(dB_{12})_i]
\Delta^2 \exp-\left\{\tilde N\mu^{-1}
\sum_{j=1}^{\tilde N}(\lambda_j-\tilde z)(A^j\cdot B_j) +\right.&
\\&\left.+ \tilde N\mu^{-1} \sum_{j=0}^{\tilde N} V(\lambda_j)
-{\tilde N}\mu^{-1}V( \tilde z)+{\tilde N}\mu^{-1}(z_0-\tilde z)((a_1 )(b_1) -
(a_2)(b_2)) \right\}&
\end{eqnarray}
The above equation includes the Vandermode for all of the eigenvalues of $z$,
 not 
just the Vandermonde of the large $N$ condensate.
In the equations above $\Delta^2$ is the Vadermonde determinant.
The sum over $j$ of $A^jB_j$ is over the color indices associated
to the eigenvalues $\lambda_1,\dots\lambda_{\tilde N}$, but not to $z_0$.
This has
been separated in the last line because we want
$a_1,b_1, a_2,b_2$ to be moduli in the matrix model,
 so they cannot be massive at the saddle point.
From this condition one of the eigenvalues of $z$ is equal to
$\tilde z$.

In the equation above it is possible to integrate $A^j,B_j$ completely.
This being a Gaussian
integral over four coordinates of mass $\lambda_j-\tilde z$
gives us a measure term equal to
$\delta= \prod_{i\neq 0}(\lambda_i-\tilde z)^{-2}$

The logarithm of the Vandermonde determinant combined with $\delta$ is then
$$\log(\Delta^2 \delta) =
\sum_{i\neq j} 2\log(\lambda_i-\lambda_j)
-2 \sum_{i\neq 0} \log(\lambda_i-\tilde z)$$
and it is the term in the effective action for the
eigenvalues $\lambda$ and $z$.
Now we want to solve for the saddle point of this setup
in the large $\tilde N$ limit. This will result in summing all of the planar
diagrams of massive fields in the above theory.
Notice that the eigenvalues $\tilde z$ and $z_0$ do not have an
 interaction between
them, because we have not integrated out the massless modes of $A,B$.
The saddle point equation for the zero components of $A$, $B$
generic make $\tilde z=z_0$ in this situation.

One sees that
the saddle point equations for the eigenvalues $\lambda_i$ are
the same as when we have the theory $U(\tilde N)\times U(0)$,
 because the contribution from $z_0$ cancels the
contribution from $z$ when they are equal. This seems accidental in 
the matrix model.
The saddle point equation for the eigenvalues $\lambda_i$ is
\begin{equation}
 \tilde N \mu^{-1} V'(\lambda_i) -
2\sum_{j\neq i} \frac 1{\lambda_i-\lambda_j} = 0
\end{equation}
Now let $w(\lambda) = \frac 1{\tilde N}\sum_{i\neq 0} \frac{1}
{\lambda-\lambda_i}$
be the resolvent of the matrix model.

The saddle point equations for $\tilde z = z_0$ are identically
equal to
\begin{equation}
 V'(\tilde z) - 2\mu w(\tilde z) +a_1b_1-a_2b_2 = 0
\end{equation}
From here, we can use the same identity that led us to the classical moduli
space to find that the quantum moduli space, as described by the matrix model,
should be given by the equation
\begin{equation}
w^2 - uv = \frac 14(V'(\tilde z) - 2\mu w(\tilde z))^2
\end{equation}
Now, we compare the matrix model prediction to the  prediction obtained  by
holomorphy arguments.
It should be the case that two different functions of $\tilde z$ are
 identical, these are as follows
\begin{equation}
\frac 14(V'(\tilde z) - 2\mu w(\tilde z))^2 =\frac 14(V'(\tilde z))^2+ f(z)
\end{equation}
where $f$ is a polynomial. We obtain from these the following
set of equations
(after changing the normalization of  $f$ to $\tilde f=-4 \mu^{-1} f$)
\begin{equation}
V'(\tilde z) w(\tilde z) -  \mu w(\tilde z)^2 = \tilde f(\tilde z)
\end{equation}
which encode the full set of loop equations of the one matrix
model \ref{eq:loop1matrix}, when we realize that $R(u)$ and $w(\tilde z)$
are really the same object.

To summarize: the quantum deformed moduli space of a supersymmetric
gauge theory  can encode loop equations for a matrix model. The above result
for a one matrix model is not new and seems redundant,
but the results in the following
sections, using the same techniques are new.

\section{Superpotential
deformations of ${\cal N} = 4$ SUSY gauge theories}

The ${\cal N}=4$ SUSY gauge theory is a supersymmetric field theory
characterized by a gauge group $G$ \footnote{The gauge group
 will be $U(N)$ for the remainder
of the paper. The results presented here can be generalized to
other gauge groups which admit a large $N$ description} and three fields
$X,Y,Z$ in the adjoint whose superpotential (up to normalization)
is given by
\begin{equation}
\tr( XYZ - XZY)
\end{equation}
This theory is a superconformal field theory
which spontaneously breaks the conformal invariance at generic points
in the moduli space,
 but if the vacuum is at the
origin in moduli space (the gauge group is unbroken) then the theory
has a vacuum with superconformal invariance.

The gauge theory with ${\cal N}=4$ SUSY has a moduli space of
conformal field theories parametrized by the gauge coupling of the theory
$g$. If we consider deformations of the superpotential, this moduli space of
${\cal N}=1$ superconformal field theories possesses a three parameter family
of deformations which give rise to superconformal field theories
\cite{LS}. These have
(up to normalization) the following form of the superpotential
\begin{equation}
W_0 = \tr( XYZ - q XZY + \frac \beta3(X^3+Y^2+Z^3))
\end{equation}
and they also respect a $\BZ_3$ symmetry which permutes
$X\to Y \to Z$.
When $\beta=0$ on top of the $U(1)_R$ symmetry, there is an additional
 non-anomalous $U(1)^2$
global symmetry of rotations of $X,Y,Z$ by phases.

It has been shown that some of these can have very
interesting brane descriptions and $AdS/CFT$
geometric duals \cite{D,BJL}, of the form
of branes on $\BC^3/(\BZ_n\times \BZ_n)$ or
$AdS_5\times S^5/(\BZ_n\times \BZ_n)$ for $\beta=0$ and $q^n=1$.
These are geometric
theories with branes on orbifold singularities, and there are three
lines of singularities in codimension two that extend to infinity
(one can think of them as the $X,Y,Z$ axis).
From here one finds
 a host of deformations of the geometry which serve to remove
 the codimension two
singularities,  leaving only codimension three
singularities behind \footnote{These codimension three singularities
are usually of the conifold type, see the examples in \cite{DF,Bcon}}.

Since these examples are noncompact Calabi-Yau manifolds,
 one has to take care in
defining the set of  allowable deformations that one can study, as the
total number of holomorphic functions describing the deformations is
infinite. The deformations of the geometry modify the superpotential of
the brane, so it is possible to analyze the geometry given a
superpotential.

For the rest of the paper we will consider only a special form  of the
general superpotential described above:

\begin{equation}
W = tr( XYZ - q XZY + V(X))
\end{equation}
Where $q^n=1$ and
\begin{equation}\label{eq:potentialmissn}
V(x) = \sum_{i\not\equiv0\mod(n)} \frac{a_i}{i} x^i
\end{equation}
is a polynomial potential.

The first condition guarantees that there is a non-anomalous
$U(1)_{YZ}$ symmetry under which $Y,Z$ have opposite charges, and is a remnant
of the $SU(4)$ R-symmetry of the ${\cal N}=4$ gauge theory.
This symmetry serves to simplify the problem.
\footnote{
One could also study by the methods presented here
a situation where
$
W = tr( XYZ - q XZY + V(X)+m(Z^2+Y^2)+\alpha Z+\alpha' Y)
$
but this example requires a lot more algebra to understand.
the one cut solution has been studied in \cite{Man}}

The second point to notice is that when the $a_i=0$ we can use 
the other $U(1)$
charge combined with the $R$
symmetry in such a way that the field $X$ has different charge than $Y,Z$
and we can also choose it in such a way that $X,Y,Z$ have all positive charge
as well as 
all of the coefficients in $V(X)$,  when we introduce
them and assign quantum numbers to the $a_i$ so that we keep the 
global symmetry. Let us call this
charge $U(1)_{R'}$

From this point of view, the deformation described above is a relevant
deformation of the gauge field theory (when considering the grading of
potential terms given by the charge), because if we consider the coefficients
as ordinary c-numbers, the $R'$ charge of the deformation is smaller than the
charge of the original superpotential.

The second condition given by equation \ref{eq:potentialmissn}
is more tricky to understand. From the string theory point of view
it amounts to the
deformations of the geometry being given exclusively by turning
fields in the twisted sector of the orbifold which have charge zero under
the $U(1)_{YZ}$ symmetry. These guarantee that the dimension of the moduli
space of a brane in the bulk remains the same as in the theory
without the deformation, namely, the deformation can be
understood geometrically and it does not produce an effective
potential on the moduli space. In what follows we will argue what form
the moduli space takes when we turn these deformations on.

\subsection{Calculating the classical
moduli space of a probe brane.}

As a supersymmetric problem, we need to solve the F-term
constraints resulting from the classical superpotential as
described above. These classical equations that determine the
vacua are given by
\begin{eqnarray}\label{eq:classrelations}
XY -q YX &=& 0\nonumber \\
ZX-q XZ &=&0\\
YZ-qZY &=& -V'(X) = - \sum _{i\not \equiv 0\mod(n)} a_i X^{i-1}\nonumber
\end{eqnarray}
and for a probe brane in the bulk
the rank of the group $N$ turns out to be equal to
$n$ \cite{D}.

In particular if we think of $X,Y,Z$ as formal symbols instead of matrices,
we obtain the result that
 finding solutions to the equations above in terms of matrices
is the
same problem as finding representations of an associative algebra over
the complex numbers,  generated by $X,Y,Z$, and subject to the relations
\ref{eq:classrelations} \cite{BJL,BL}, which are exactly of dimension
$n$. We are particularly interested in irreducible representations of
the above algebra of dimension $n$, because this is the rank associated to a
brane in the bulk. The irreducibility implies that the gauge group is 
broken to $U(1)$ in the infrared, which is the gauge group
associated to a single brane in the bulk.

Finding these representations will let us calculate the classical moduli
space of the theory and recover the Calabi-Yau geometry.
The holomorphic
 functions on the Calabi-Yau geometry will be given by functions
which are in the center of the above algebra
\cite{BJL,BL,Brev}; and the relations between
these functions will be the relations that follow by virtue of
imposing the equations \ref{eq:classrelations}.
In an irreducible
representation all of these variables in the center can be shown
to be proportional to the identity due to Schur's lemma. Hence
these are essentially gauge invariant, as they can be recovered completely
from their trace. From this point of view, the relations in the center
of the algebra
are relations in the classical chiral ring of a one probe brane
system.

We need to solve the problem of finding the set of irreducible
representations of dimension $n$ of the above algebra and the center
of the algebra. To do this we need to solve for the F-terms above in
dimension $n$.

Solving the first equation is easy if we introduce the clock and shift
matrices
\begin{equation}
P = \begin{pmatrix} 1&&&&\\
&q&&&\\
&&q^2&&\\
&&&\ddots&\\
&&&&q^{n-1}
 \end{pmatrix}, Q = \begin{pmatrix}0&1&0&\dots&0\\
0&0&1&\dots&0\\
\vdots&\vdots & \vdots &\ddots&0\\
0&0 & 0& \dots&   1\\
1&0&0&\dots&0\end{pmatrix}
\end{equation}
which satisfy $QP = q PQ$, $Q^n= P^n=1$. These two matrices are linearly
independent and generate the matrix ring of $n\times n$ matrices.
A general $n\times n$ matrix $M$ can be written in a unique form
 as a linear combination
$M = \sum_{i,j=0}^{n-1} a_{ij} P^iQ^j$.

One can show that the first equation in \ref{eq:classrelations}
 is solved (generically)
up to equivalence by $X = x P$ and $Y= y Q$. This is the case when
$Y^n =y^n\neq 0$ and $X^n=x^n\neq 0$.
To solve the second equation, the most general solution has to be of the
form
\begin{equation}
Z = z Q^{-1}P^{-1}  + \sum_{i=1}^{n-1} z_iQ^{-1} P^{i-1}
\end{equation}
If we now substitute this expression  in the third equation, we obtain that
\begin{equation}\label{eq:special}
\sum  y z_i P^{i-1} (1-q^{i}) = - \sum a_i x^{i-1} P^{i-1}
\end{equation}
Since the powers $P^i$ are linearly independent, we can solve for the
$z_i$ as follows
\begin{equation}
z_i = -a_i y^{-1} x^{i-1}/(1-q^{i})
\end{equation}

Thus we find that the generic irreducible representations of
dimension $n$  are parametrized by three variables $x,y,z$.
However, one can show that various values of $x,y,z$ can
correspond to the same irreducible representation. The equivalent
classes can be distinguished by evaluating the coordinates in the
center of the algebra. Notice that the equation \ref{eq:special}
can only be solved if no term proportional to $P^{-1}$ appears
on the right hand side of the equation. This is exactly the
restriction that was imposed on the form of the superpotential.

We have found above a three-parameter family of irreducible
representations. However, gauge invariance makes it possible
to change $x\to qx $ with some other modifications on $y$ and
$z$, so the variables above describe a cover of the moduli
space. We need to evaluate gauge invariant combinations of
$x,y,z$.

It is easy to show that $u= X^n= x^n$ is in the
center of the algebra. From the representation above it is also
easy to show
that $v=Y^n=y^n$ is proportional to the identity in the representation.
Since $Y,Z$ appear
essentially
on the same footing in the equations \ref{eq:classrelations}, then it follows
that $w= Z^n$ is also in the center; although it is difficult to
write an expression for $Z^n$ in terms of the $a_i$, $x,y,z$.
 Finally, there is another variable in the center, which we will call
$t$ and  whose form is given by $t= XYZ+f(X)$, where $f$ a polynomial.
$XYZ$ can be readily calculated in the representation above to
give
\begin{equation}
XYZ = xyz+ \sum_{i=1}^{n-1}  z_iy x P^i
= xyz - \sum a_i(1-q^i)^{-1} x^{i} P^i = t - f(X)
\end{equation}
so that $f(X) = \sum a_i(1-q^i)^{-1} X^i$, and $t=xyz$.

We see that in this particular matrix representation $u,v,t$ are easy
to calculate, and $w$ is hard to calculate. Also, these are four
variables determined algebraically
by three parameters $x,y,z$, so they must satisfy some
algebraic relation.

In the particular case where
$V'(X)=0$, one easily finds that $w= Z^n = z^n$, and then the
variables $u,v,w,t$ satisfy the following equation
\begin{equation}
u v w = t^n
\end{equation}
With the addition of $V$, the relation above is
deformed.
The importance of the calculation done above was to find  the
expression that described the variable $t$.

The classically deformed relation above is relatively
easy to calculate if we
choose to write $Z$ in a slightly different form, so that $w$ can
be read more easily
\begin{equation}
Z = \begin{pmatrix}0 &&&&\zeta_{0}\\
\zeta_1&0 &&&\\
&\zeta_2&0&&\\
& &\ddots&\ddots&\\
&&&\zeta_{n-1}&0
\end{pmatrix}
\end{equation}
Then we find that $Z^n = \prod \zeta_i$, and we can write equations for
the $\zeta_i$ in terms of $t$. Namely
\begin{equation}
XYZ =xy \begin{pmatrix}\zeta_{0} &&&\\
&\zeta_1 &&\\
&& \ddots &\\
&&&\zeta_n
\end{pmatrix} = t - f(X)
\end{equation}
which reduce to $xy\zeta_i = t-f(q^ix)$. Taking the product of all
of these equations we find
\begin{equation}
x^n y^n\prod \zeta_i = uvw
= \prod (t-f(q^i x))
\end{equation}
Now, because the right hand side is invariant when we change
$x \to q x$, the right hand side only depends on $u=x^n$ and $t$.
 So we find that the deformed geometry corresponding to the
 change in the
superpotential generated by $V$ is the following hypersurface in
four complex variables $u,v,w,t$
\begin{equation}
u v w = Q(t,u)
\end{equation}
where $Q$ is a polynomial in $t$ and $u$ of order $n$ in $t$
and of the same  order as $V'$ in $u$.

One can also notice that the polynomial $Q(t,u)$ is not a generic
polynomial in $t,u$ of the given order, but it depends on fewer
parameters (only the variables $a_i$).

When this happens, the geometry described above has some
special properties that make it non-generic.
In the above case, one can expect that the non-genericity of $Q(t,u)$ is
related to the existence of singularities for the curve in the $u,t$
variables described by $Q(t,u)=0$.
The singularities of the CY geometry can be obtained by demanding that
the partial derivative with respect to the variables $u,v,w,t$ of
the hypersurface equation vanish.

The derivatives with respect to $v,w$ give rise to the following equations
\begin{equation}
uv = uw =0
\end{equation}
These can be solved for general $u\neq 0 $ by
$v=w=0$. Under these conditions the CY has singularities at some values of
$t,u$ for which the curve $Q(t,u)=0$ is singular.
These singularities are in general of the conifold type.
 These are easy to find: just look for
repeated roots of $t$. These are values of $x$ such that
$f(q^r x) =f(x)$ for some $r\neq 0$.

A second set of singularities is produced by
setting $u=0$. In this case it is easy to show that $Q(t,0)=t^n$, and
for $n\neq 1$ the derivative of $Q$ with respect to $t$ vanishes at $t=0$.
This produces a curve of singularities characterized by the equation
$vw = Q_u(t,u)|_{u=0}$.  This curve
degenerates to
two of the three lines of singularities that are present in the orbifold
geometry when $Q_u(t,0)=0$. This is a curve of
$A_{n-1}$ singularities.

The singularities which will be important for us are the isolated
singularities that we described first, and which are of the conifold type.
 At a conifold singularity there are usually
two types of fractional branes present (even in the case where the
singularities seem to have discrete torsion \cite{Bcon}),
 which wrap the local cycle of zero
geometric size with the two possible different orientations.

The fractional branes in the above situation are obtained by taking a brane
in the bulk to a singularity. The brane representation then becomes
reducible and can be written as a direct sum of two sub-representations
$\lim R = R^+_\alpha, R^-_{\alpha}$.
Each of these sub-representations behaves in the classical
low energy theory exactly as in the conifold field theory described in
\cite{KW}, so all of the adjoint fields are massive when we
consider these representations.

Let us write these representations for generic $q$.
The idea is that the algebra of the F-terms
 will be a slightly modified $SU_q(2)$ algebra, so the solutions will look
essentially like representations of $SU(2)$.
We can choose these to be of dimension $k$, where $Y,Z$ are like ladder
operators.
\begin{equation}
Y = \begin{pmatrix}0&y_1&0&\dots\\
0&0&y_2&\dots\\
\vdots&\vdots&\ddots&\ddots
\end{pmatrix}
\quad Z = \begin{pmatrix}
0&0&0&\dots\\
\zeta_1&0&0&\dots\\
0& \zeta_2&0&\dots\\
\vdots&\vdots&\ddots&\ddots
\end{pmatrix}
\end{equation}
The commutation relations with $X$ force $X$ to have the following form
\begin{equation}
X= x\diag(1,q,q^2, \dots, q^{k-1})
\end{equation}
Finally, the commutation relations of $Y,Z$ give linear relations that
overdetermine the
$y_i\zeta_i$. It is even possible to choose a gauge where $y_i=1$, so that
only the $\zeta_i$ are variables.
Consistency of the equations gives a finite number of roots $x$
that solve the problem, and the number of different
roots is equal to the degree of $V'(x)$. Irreducibility requires that
all of the $\zeta_i$ are different from zero.
For $q^n=1$,
if $k\geq n$ one can show that the above representations are
reducible, because the $\zeta_i$ end up being periodic and a zero invariably
shows up.

The ones that survive for $q^n=1$ are exactly the ones that are associated to
fractional branes in the geometry, and have rank less than $n$.
If we evaluate $t$ as defined in the bulk
for the fractional brane representations, we find that $t=f(q^{k-1}x)$; and
the condition on $x$ to be a ``root''  are that $f(x)=f(q^{k}x)$.

Notice that there is no gauge
symmetry sending $x\to qx$ for the fractional
branes. This means that a fractional brane contributes to traces
of the form $\tr(X^k)$ for all $k$. These gauge invariant variables are not
traces of elements of the center of the algebra.

Also, at roots of unity for $q$, of the infinite number of representations
of the slightly deformed $SU_q(2)$ we have for general $q$,
 only finitely many survive, and
these are exactly the representations of dimension less than $n$.
This is the property
that will give a finite number of cuts later
in the matrix model.

The fractional branes can lead to isolated vacua of a $U(N)$
theory for arbitrary $N$. These are exactly like the representations
of $SU(2)$ that are familiar from the ${\cal N}=1^*$ system.
There are also fractional branes at the curve of
singularities. These have a moduli space of vacua which is given by a
connected $n$-fold cover of the curve. Given $n$ branes in the curve of
singularities one can move them so that they sit on top of each other
in the CY geometry, but that each of them is in a different leaf of the
cover. This configuration is produced by taking a brane in the bulk
to the line of singularities.

The description of the classical vacua of the
theory for $U(N)$ is given by splitting into irreducible representations
of the algebra.
$$
R = \oplus N_iR_i
$$
and lead to a theory which in the infrared has a $\prod U(N_i)$ gauge
symmetry, with $N= \sum N_i\dim(R_i)$.
 If the $R_i$ are branes in the bulk
or in the curve of singularities,
then one has a moduli space of vacua. Even if the $R_i$ are fractional branes,
$R^{\pm}_\alpha$, there can be a moduli space for these branes if they
 are accompanied by $R^{mp}_{\alpha}$.

The branes in the bulk behave in the low energy physics like ${\cal N}=4$
field theory, so we can take them to a generic point in moduli space where
they are away from the singularities, and also remove  paired fractional
branes to a generic point in moduli space in the bulk. After this is done,
there are some fractional branes that can not be
removed from the singularities by moving inside the classical moduli space.
The number of branes stuck at each isolated singularity is
constant in the moduli space (this is a classical statement).

In this form we realize that the description of the moduli space of the
$U(N)$ theory comes in branches of different dimension, and the branches are
characterized by how we choose to distribute fractional branes at the
isolated singularities.

Among all of these possibilities, one can always find solutions where
there are only fractional  branes stuck at the isolated singularities.
This is because there is at least one fractional brane with
$\dim(R_\alpha)=1$. In general we can have various
isolated vacua where the
fractional branes are distributed in some form at the singularities.
These isolated vacua are exactly of the type that one believes one
understands, because in the infrared one has pure gauge theory degrees of 
freedom. The $SU(N_i)$ gauge groups should 
confine and the low energy theory reduce to $\prod U(1)$, just like 
in the ${\cal N}=2$ SYM softly broken to ${\cal N}=1$.

Geometrically, if we place fractional branes
at conifold singularities, these are
expected to take the geometry through a geometric transition and
deform the singularity away, if they have the same 
behavior as described in \cite{KS,GV}. This produces a new geometry where
the quantum effects of the branes are taken into account.  We will
 argue later, based on holomorphy that this should
correspond to having a  quantum
corrected curve $\tilde Q(t,u)$, with a more generic polynomial of
degree $n$ in $t$ and $s$ in $u$.

We can now ask the question: how will we be able to see this
geometry? The intuition is that if we use a probe brane, then the
brane will have as it's moduli space the deformed geometry. This intuition
is valid so long as we can ignore the back-reaction of the branes at
the singularities to the probe.           From the string theory point
 of view, this will be a large $N_i$ limit. However, we will see that we
can predict using matrix models
that there is no back reaction even at finite $N_i$.

The issue we should now consider, is that for every distribution of fractional
branes we get a different deformation. In the string geometry this is because
the geometric transition exchanges the algebraic cycle that
the fractional branes are
wrapping, to an $S^3$ with flux.
The flux is given by the number of branes
wrapping the cycle, and the flux tries to make the volume of the associated
$S^3$ bigger in order to reduce the energy in the flux. Thus, we have to
consider each vacuum configuration independently.

In particular, the description of the quantum deformed
moduli space of a probe brane has to be different
for each branch of vacua. This is very similar to the discussion
given in \cite{CSW}, where the order parameters that distinguish the
vacua are equations which are only satisfied on some branches of the theory
 and not others.
If we stop to think for  a minute, the discussion we have done so far
even at the classical level has this property. The equation of the CY
geometry in the gauge theory assumes that we have split the classical
representation of the vacua into irreducible representations, and the
equation is only valid representation by representation. This is, we have
made a choice of {\em eigenvalues} (we can think of block diagonal matrices as
a generized form of eigenvalue), and the equations are valid eigenvalue by
eigenvalue.

However, the eigenvalues themselves are not  gauge invariant, as we
have to account for permutations of the eigenvalues. Only the symmetric
functions of the eigenvalues are truly gauge invariant. However, the CY
equations described above are not valid when we sum over eigenvalues.
They are only valid eigenvalue by eigenvalue. To classically obtain the
eigenvalues from the symmetric functions amounts to solving a polynomial
and choosing a labeling for the roots. This extra label is the non-gauge
invariant information that tells us that classically we are not studying
{\em universal equations} on the chiral ring of the gauge theory,
 but rather equations in the chiral ring that are valid only on a particular
 branch of the theory.

This subtle point is very important. It means that the {\em non-universal}
equations that determine the
quantum moduli space does not only receive contributions from instantons,
but it can be corrected by other strong dynamics effects, such as gaugino
condensation.

\subsection{The quantum moduli space}

We have exhausted the discussion of the classical moduli space of the theory.
Now we want to study the moduli space of one probe in the presence of
fractional branes at the singularities, whereas in the previous section we
discussed essentially the moduli space of one probe alone, plus the classical
description of the branches of the gauge theory.
We want to understand the quantum effects of the fractional branes
on the geometry, and we will use a probe brane to test the shape of the
geometry.

Before we study these quantum deformations, one should
make a field theory argument for the moduli space
not being lifted, namely, for the absence of a potential on the moduli
space.

The argument one can make is semiclassical. If one examines the theory at
infinity in moduli space, all the massive fields that connect the brane
in the bulk to the fractional branes
have a mass much higher than any confining scale. This means
one should be able to integrate them out completely and reduce the problem
to low energy effective field theory. The effective field theory on the
probe brane is essentially ${\cal N}=4$ SUSY gauge theory locally, and we
do not expect that to change too much.
The effective
superpotential of the brane in the bulk should decay at infinity, because we
are in a semiclassical regime. Now, the vev of the superpotential is
controlled by holomorphy, and it vanishes at infinity. If the moduli
 space is lifted, then the superpotential must be a holomorphic
function on the moduli space,
and it should have poles or vanish exactly.
The only possibility of having poles is if there are codimension one
singularities on the classical moduli space.
However, the moduli space described above
is a Calabi-Yau geometry, which cannot have singularities
in codimension one. This means that there is no effective
potential generated on the moduli space, and the moduli space is not lifted.
This does not prevent the moduli space from getting corrected. Indeed, our
argument in the previous 
subsection suggests that it does, and when it does, the constraints
that the deformed moduli space satisfies  are controlled by
holomorphy. This argument should be generic for moduli spaces of bulk-branes
in a geometrically engineered
theory, as in that case the moduli space of vacua of a brane in the
bulk is always a CY geometry.

To check the possible shape of the geometry we need to consider the
$U(1)_{R'}$ charge we defined. It is clear that when we keep the charges of
the coefficients of $V$ into account, the equation describing the moduli
space is homogeneous for the
$U(1)_{R'}$ charge. Since this is a non-anomalous symmetry of
the gauge theory, this remains true in the quantum moduli space.
A (partial) gaugino condensate $S_i$ will have the same charge as the
variable $t$ (this is the charge of the superpotential term).

However, $u$ can have a charge that is not commensurate with $S$, because
we had some freedom in the definition of the $U(1)_{R'}$ charge. Also
$v,w$ are charged
with respect to the $U(1)_{YZ}$ symmetry, so they must appear together.

Holomorphy implies that all of the couplings, and $S$ appear with positive
powers in the deformed equation. Also, the moduli space is not expected to
develop new branches, so it must also be polynomial in $u,v,w,t$, and it
should reduce to  the classical moduli space if we set the $S_i$
to zero. The $R'$ charge of the $S$ allow us to replace monomials where a
$t$ appears with an $S$; the coefficients depend on the branch
of moduli space one is studying, but we can not change the coefficient of
$vw$ nor include higher powers of $vw$. This results in a deformed CY
geometry where the equation that describes the geometry has the form
\begin{equation}
uvw = Q(t,u)+\delta(t,u)= \tilde Q(t,u)
\end{equation}
where we have deformed the curve $Q(t,u)$ to $\tilde Q(t,u)$, making sure that
the $R'$ charge is conserved when we keep the information of the charge of 
the gaugino condensate.

Now, let us take an example to illustrate the shape of the deformations,
where we set
$n=2$ and $s=3$.
In this case $V(x) = a_1 x + a_3/3 x^3$,
$2 f(x) = a_1x +a_3 x^3 = - 2 f(-x) =x( a_1 + a_3 u)$, and the
curve is given by
\begin{equation}
Q(t,u) = (t-x/2(a_1+a_3 u))(t+x/2(a_1+a_3 u))
= t^2 - u (a_1+a_3 u)^2/4
\end{equation}
The singularity of $Q(t,u)$ occurs at $u= -a_1/a_3, t=0$. In
principle the allowed deformation of the polynomial are
given by
 $\delta(t,u) = r_0tu+ r_1 t+ \beta_1 u +\beta_0$.
The other coefficients, which are of higher degree,  are
determined uniquely by $a_1,a_3$, the bare parameters of the
superpotential. $r_0$ and $r_1$ can be eliminated by a linear
change in variables in $t$, so we drop them. In general we can
always eliminate any term proportional to $t^{n-1}$, so $\tilde
Q(t,u) = t^n+ t^{n-2} O_1(u)+t^{n-3} O_2(u)+\dots$.

 Moreover,
 we have not placed branes at the curve of singularities located
at $u=0$, so we are not allowed to deform these singularities
away. This forbids us from adding the term $\beta_0$. In the end
the only allowed deformation is given just by $\beta_1 u$. This is
a one parameter deformation of the geometry. This is expected
because we have only one conifold singularity to resolve. The
counting of parameters that deform the geometry should be equal to
the maximum number of conifold singularities that the geometry can
have. When these conifolds collide they can generate more complicated
singularities, but the counting of normalizable deformations should remain
invariant.

In the general case the polynomial $Q(t,u)$ has no term added
which does not contain a power of $u$, because we do not place
fractional branes at the curve of $A_{n-1}$ singularities,
only at isolated singularities. This gives rise to constraints on
the coefficients of the curve.
 The reader can verify that the deformed CY
geometry still has a curve of singularities at $t=u=0$ with the
above prescription of not including any deformation with just a
different power of $t$ which does not involve $u$ as well.
The general idea is that placing
fractional branes at a given singularity only resolves that
singularity, and leaves the other singularities intact. In fact this argument
has been used in various papers \cite{CV,CSW} 
to study factorization properties of Seiberg-Witten curves.
The geometrical idea is that one
stabilizes the 3-cycles by flux,
 and in the absence of flux, the cycle shrinks to zero size.
At least
from this point of view one can characterize the configurations
with no branes at some singularities: these are solutions where
the curve $\tilde Q(t,u)$ preserves these singularities, and is
therefore singular.

We have presented a holomorphy argument to determine the
deformed geometry as seen by a probe in the bulk, when we have placed
fractional branes at the conifold singularities. The reader should be warned
that the arguments given above assume that there is a way to define the
variables $u,v,w,t$ for a given configuration. Even classically, this
depends on ``removing'' the information of the fractional branes to extract
these variables. In the quantum theory this procedure gets corrections, so
it becomes very tricky to define these in a sensible way. Even the
partial gaugino condensates need a good definition in the field
theory \cite{CDSW}.
 The underlying assumption of these ideas  is that all of
this can be done in a systematic way, and that there
is a procedure to determine these variables
in terms of vevs of elements of the chiral
ring of the gauge theory. This has to be done 
in each particular branch of the theory separately, and
with only one probe brane in the bulk.

We will later show
that as in the example of the previous section,
the deformed geometry encodes relations for some
generalized resolvents in a matrix model.
This result can be used to find recursion relations which 'solve' the
matrix model in the sense expressed in the
introduction.

\subsection{The chiral ring of the gauge field theory}

We have argued what the shape of the deformed moduli space of vacua is, as it
pertains to a single bulk D-brane. In some sense, this solves the structure
of the chiral ring for a single brane in the bulk.

As we have argued, the description of the classical moduli space depends
on singling out eigenvalues and performing various operations with classical
matrices. In the quantum theory all of these statements should be interpreted
as operator equations in the chiral ring, so it is important to classify the
chiral ring of the full gauge theory, not just of the ``bulk'' brane.

The description is very similar to the one matrix model, except that now
we have three matrices. Also, we should regard the rank of the matrices as an
unknown, as we can have arbitrary fractional branes at the various
singularities.

The chiral ring is made of gauge invariant operators,
and factorization is guaranteed by the cluster
decomposition principle when we study chiral rings vevs in a
{\em supersymmetric vacuum}. It is under these conditions that we actually
believe that it is possible to relate the supersymmetric gauge theory to
the associated matrix model.

Following the setup of \cite{CDSW}, the
chiral ring is generated by traces of the following form
\begin{eqnarray}
tr(f(X,Y,Z))\\
tr(f(X,Y,Z)W_\alpha) \\
tr(f(X,Y,Z)W_\alpha W^\alpha)
\end{eqnarray}
where $f$ is a polynomial in $X,Y,Z$.
We have to remember that $X,Y,Z$ are matrices, so the ordering inside
$f$ of the different letters $X,Y,Z$ matters. The ordering of $W_\alpha$
is not important, because adjoint action of $W_\alpha$
on a set of letters is a total derivative. This allows us to choose an
ordering where all the $W$'s are together.

Also, no more than two $W_\alpha$ per trace are allowed because these terms
are a total chiral  $\bar D$ derivative \cite{CDSW}.

The chiral ring generators as presented above are independent of $N$, and
of the specific details in the superpotential.
The generalized Konishi anomaly equations for variations of the fields
will depend on the specific form of the superpotential, and these are
the ones that can be related to a matrix model.

Classically, the relations in the chiral ring are
given by solving the F-term constraints
of the superpotential. These can be written in a gauge invariant way in
the following form
\begin{equation}
\tr(\frac{\partial W}{\partial\phi} f(X,Y,Z))=0
\end{equation}
for $f$ any word in the fields, and $\phi$ any of
$X,Y,Z$. These are essentially the same equations as $W'=0$, since it implies
that no matter what matrix we  build out of $X,Y,Z$, the trace is
zero.

If we include quantum corrections, the equations of motion take the form
\begin{equation}
\tr(W'f ) = \hbox {Quantum corrections}
\end{equation}
The quantum corrections involve the (partial) gaugino condensate (s)
of the (fractional branes) gauge theory,
and we should interpret the above equation
as a part of a generalized Konishi anomaly, where $f$ can include
the $W_\alpha W^\alpha$.

A few points are worth noticing: when we fix the superpotential as in the
 previous subsection, branes in the bulk do not contribute
classically to one point functions of the form
\begin{equation}
tr(X^k), tr(Y^k), tr(Z^k)
\end{equation}
for $k\neq 0\mod (n)$, nor to any word containing a gaugino field
\footnote{Classically the gaugino field has to be zero for Lorentz
invariance, and this implies that the  gaugino condensate classically
is also zero}.
This can be seen because the classical solutions of branes in the bulk
give contributions to the above as sums of the form
$$
\sum_{i} q^{ik}=0
$$

In the orbifold theory, all of these words correspond
to twisted fields. The classical string theory
statement is that a brane in the bulk
does not carry charge in the twisted sector of the orbifold,
so it can not generate vevs for twisted fields.

Fractional branes, on the other hand, do couple to the twisted sector.
It follows
that they generate vevs for twisted sector words. We should read this
statement in the following way: classically, there is information
that depends only on the fractional brane structure.

Moreover, we are interested in isolating the fractional branes that are
stuck to the codimension three singularity, in a situation where there is no
moduli space of vacua. In the gauge theory the low energy physics reduces
to pure gauge theory $\prod U(N_i)$, and we can argue that the
$SU(N_i)$ confine and give rise to some partial gaugino condensate.

Guaranteeing that there are no branes in the codimension two singularities
of the geometry is a simple matter. These contribute to
the vevs $\tr(Y^k)$ and $\tr(Z^k)$ for all $k$. Also, branes in the bulk can
contribute to $\tr(Y^k), \tr(Z^k), \tr(X^k)$ for $k$ multiples of $n$.

This means that classically these can be considered variables that we can
set to any value we want. Notice that $\tr(X^n)$ and $\tr(X^{2n})$ are
uncorrelated if we have more than one brane in the bulk. In particular, we
 can always consider
enough branes in the bulk so that these variables are allowed to have any
value.

The same is not true for the other powers of $\tr(X^k)$. There are only a
finite number of values of $X$ that are allowed for a fractional brane, so
 these variables are related to each other in the classical theory;
there will be recursion relations that determine these
variables. The data necessary to seed this information to the classical
recursion
relations is exactly
how many fractional branes there are at each singularity, and which
cycle they are wrapping.

One can wonder if in the quantum theory the vevs of
$\tr(X^k)$ and $\tr(X^kW_\alpha W^\alpha)$ for $k$ not a
multiple of $n$ can get
contributions from branes in the bulk.
We will argue that for the second ones, which have a simple matrix model
interpretation this is not so.
A hand waving argument would say that the gauge groups
that have gaugino condensation are just the fractional  branes and not
the branes in the bulk, so they contribute to these one point
functions. In principle, if these were classical matrices,
 one could make the same statement about
any trace containing $W^\alpha W_\alpha$. However, the
quantum vacuum is not given by classical matrices,
and one should consider only twisted operators.

At infinity the branes in the bulk
should be semiclassical and decouple, so
 their effective gaugino condensate should vanish.
Thus the contribution to $\tr(X^kW^2)$ should be
suppressed by a large  mass scale. At infinity, their contribution is
zero, and $\tr(X^kW^2)$, being holomorphic on the moduli space, is constant
by the same argument that made the effective superpotential on the moduli
space equal to zero. This suggests that these one point functions
 are determined by the fractional branes alone.

Since these are related to $\tr(X^k)$, and classically the twisted
$\tr(X^k)$ satisfy recursion relations (they are determined
alone by the fractional branes), it suggests that the same is true for
$\tr(X^kW^2)$.
 The description above is suggestive. It is not meant to
be a formal argument.
We will use matrix models to make this intuition precise.

\subsection{Relations to a matrix model}

We have found that we can write the most general form of the
deformation of the Calabi-Yau geometry which is compatible with
holomorphy. This information in the gauge theory encodes non-perturbative
dynamics that results from gaugino condensation
on the gauge group of the fractional
branes. All of these effects should be
 computable from a matrix model.
What we have described is the moduli space of a
probe brane. The matrix model
calculation proceeds in the same way we worked in
section \ref{sec:1matrix}. We isolate the block of matrices associated
to the probe brane. This is done by (perturbatively)
identifying a set of eigenvalues for $X$ ( an $n\times n$
block matrix) to correspond to a bulk brane, where the eigenvalues
 are related to each other
in such a way that they produce massless field in the classical field
theory.

In the end, we have to
consider the three matrix model whose potential is given by
\begin{equation}\label{eq:potential}
W(X,Y,Z) = N\mu^{-1} [\tr( XYZ-qXZY) + \tr V(X)]
\end{equation}
with  $n$ correlated eigenvalues which correspond to a single
brane in the bulk, and take the large $N$ limit.

We are after the effective potential for the set of block
matrices in the bulk.
Notice that we can choose the gauge where
$X$ is diagonal, and we can isolate the block
of matrices which correspond to the eigenvalues in the bulk.

The idea now is to integrate the off-diagonal elements that connect
the eigenvalues of the bulk to the eigenvalues of the fractional
branes. The procedure is to
integrate these fields in the matrix model, where we take a large $N$ limit
for the matrices and we will obtain
an effective potential for
$X_{bulk},Y_{bulk}, Z_{bulk}$ with the effects of the
fractional branes included.

This should be seen in the same philosophy of the proof of the matrix
model conjecture as given in \cite{DGLVZ}. The new ingredient is that
perturbatively massless fields can not be integrated out. We also have to
remember that since the classical moduli space has singularities, at each
singularity new massless fields appear. The $n\times n$ block of matrices
contains all of these fields, so we need to keep them unintegrated to
understand what happened to the singularities.

To compute the quantum corrected moduli space then we need to vary the
effective potential for $X_{bulk},Y_{bulk}, Z_{bulk}$ as $n\times n$ matrices
and set the variations  equal to zero.
We then need to evaluate what are the quantum corrections
to the form of $u,v,w,t$ for the bulk matrices
and find the new relations between them.
This is technically a little bit
more elaborate than the situation in section \ref{sec:1matrix}.
 It is useful to keep in mind  that the brane in the bulk is auxiliary:
it is giving us information about the matrix model of
the branes stuck at the singularities. We want to know exactly
what information it is encoding.

To do the first part, we divide the matrices as follows
\begin{equation}
X = \begin{pmatrix}
X_b & 0\\
0&\framebox{ $\begin{matrix}
\lambda_1 &&&\\
&\lambda_2&&\\
&&\lambda_3&\\
&&&\ddots
\end{matrix}$}
\end{pmatrix},
Y = \begin{pmatrix} Y_{b} & Y_{bf}\\
Y_{fb}& Y_f
\end{pmatrix}
, Z = \begin{pmatrix} Z_{b} & Z_{bf}\\
Z_{fb}& Z_f
\end{pmatrix}
\end{equation}
where we have taken care to write the matrix $X$ in a block
diagonal form, and where the eigenvalues attached to the singularities
are singled out. We keep the matrix of the bulk $X_b$ in
general form. In the matrix model, with $X_b$ unintegrated,
requiring that the $X_{bf}=0$ is a gauge choice.
This introduces ghosts \cite{DGKV}, whose measure will give the Vandermonde
determinant.
The subindices are labeled as a mnemonic: $b$ for bulk, $f$ for
fractional.
 Integrating over the
orbit of the gauge group introduces a factor of the volume of the
orbit, which is proportional to the Vandermode determinant.
This includes both the eigenvalues at the singularities and the
block matrix $X_b$, even if we choose not to diagonalize $X_b$, there is
still a contribution from how the eigenvalues $\lambda$ `repel'
the block $X_b$.

When we take the logarithm of this measure term, it
gives us a contribution to the effective potential
which is given by
\begin{equation}
2\sum_i\tr_b\log( X_b-\lambda_i) + \sum_{i\neq j} \log(\lambda_i-\lambda_j)
\end{equation}
We have only written the terms involving $X_b$, as this is
the information that relates to the brane in the bulk
\footnote{The saddle point equations for the eigenvalues $\lambda$
are discussed in the appendix.}.

 We now set
$X_{bf}=0$ everywhere in $W(X,Y,Z)$ and remember to keep track of
the Vandermonde determinant.

It is easy to see that after this substitution $Y_{bf}, Y_{fb}$
and $Z_{fb}, Z_{bf}$ appear quadratically in the equation, and can
be integrated completely, separating the matrices from the bulk
from the matrices associated to the fractional branes, except for
quantum effects. These two off-diagonal integrations contribute
another effective term to the potential of $X_b$ alone, which is
given by
\begin{equation}
-\sum_i \tr_b(\log(X_b-q\lambda_i)- \sum_i\log(X_b-q^{-1}\lambda_i))
\end{equation}
Again, we obtain a formal power series in $X_b$ which encodes loop operators of
the fractional brane matrices.

The effective potential for $X_b,Y_b,Z_b$ in the matrix model is then given
by
\begin{eqnarray}
W &= N\mu^{-1}&( \tr(X_bY_bZ_b-qX_bZ_bY_b)
+V(X_b))\\&& -\sum_i \left[
2log(X_b-\lambda_i)-\log(X_b-q\lambda_i)-
\sum_i\log(X_b-q^{-1}\lambda_i)\right]
\end{eqnarray}
Now, we have to assume that the distribution of the $\lambda_i$ is
known, so that we can think of the terms involving logs as a
formal function of $X_b$. Notice that this formal function has a
power series expansion in $X_b$ which has the same characteristics
as $V(X_b)$: no powers of $X_b$ which are multiples of  $n$ appear.
The effect is the same as if we would have done a change in the
potential $$V\to \tilde V = V - \frac{\mu}N \sum_i \left[
2log(X-\lambda_i)-\log(X-q\lambda_i)-
\sum_i\log(X-q^{-1}\lambda_i)\right]  $$ so the moduli space has
essentially the same form as we discussed in the previous section,
 except that the new
$\tilde Q$
now has information about the distribution of eigenvalues
$\lambda_i$. Notice that the formal form of the solution gives us
a curve of the following form
$$
uvw = Q_{matrix}(t,u)
$$
where in principle $Q_{matrix}(t,u)$ is a 
formal power series in $u^{-1}$.

The new calculation of $\tilde V(X)'$ gives the following result
\begin{eqnarray}
\tilde V'(X)  &= V'(X)+& \frac{\mu}N\left[\sum_i 2\frac
1{X-\lambda_i} - \frac1{X-q\lambda_i}
-\frac1{X-q^{-1}\lambda_i}\right]\\
&= V'(X)+& \frac{\mu}N \sum_i \sum_{j=1}^{n-1} (2-q^j-q^{-j})
\frac{X^{n-j-1} \lambda_i^{j}}{x^n-\lambda_i^n}
\end{eqnarray}
where $u=x^n=X^n$ is still in the center of the modified  algebra.

We can define the following $n-1$ generating functions of moments of
the eigenvalues $\lambda_i$,
\begin{equation}
R_j (\gamma) =  \frac 1 N \sum_i\frac{\lambda_i^{j}}{\gamma-\lambda_i^n}
= \sum_{k=0}^\infty\frac {< \lambda^{nk+j}>}{\gamma^{k+1}}
\end{equation}
and write the (matrix version of the) quantum moduli space
in terms of these generating
functions.

The quantum version of $f(x)$ is
\begin{equation}
\tilde f(X) = f(X) + \sum_{j=1}^{n-1} (1-q^j)X^{n-j} R_j(u)
\end{equation}

Now we can compare both descriptions of the quantum moduli space:
the one we obtained using holomorphy arguments, and the one where we assumed
that we had the knowledge of the $R_j(u)$. The main assumption in this paper 
is that
both constructions are giving exactly the same moduli space. Indeed, the
one difference
between $\tilde Q(t,u)$ and $\tilde Q_{matrix}(t,u)$ is that the first one
 is a polynomial, and the second one is only a formal power series.

The equality of the
moduli spaces means that we can write equations that the $R_j(u)$
must satisfy: one for each power of $t$, so that the matrix model
gives also a polynomial. There are $n-1$ unknown
functions $R_j(u)$, and there are $n-1$ non-trivial relations, one for each
power of $t$ starting at $t^{n-2}$. The equation for $t^{n-1}$ is
satisfied trivially because the product defining $Q$ is over all
the values of $\tilde f(q^iX)$, and $\tilde f$ does not contain
integral powers of $X^n$.
This means the solution is algebraic.

The other equations are given by expanding both forms of $\tilde
Q(t,u)$ and comparing coefficients. We will call these the Quantum
relations of the matrix model.  The first two are explicitly
written bellow:

\begin{eqnarray}\label{eq:quantum}
O_1(u) &= & \sum_{i<j}\tilde f(q^i x)\tilde f(q^j x)\\
-O_2(u) &=& \sum_{i<j<k}\tilde f(q^i x)\tilde f(q^j x)\tilde f(q^k
x)
\end{eqnarray}
The right hand side is given by explicit power series of $u^{-1}$ in
terms of the one point functions $<\lambda^k>$, and all of the
coefficients except the first few (the ones that have positive powers of $u$) 
vanish. This gives us
constraints that the one point functions must satisfy, and one can
notice that they give rise to
 recursion relations for the moments of the eigenvalues
$<\lambda^k>$,
so that all but a finite number of them are determined in terms of
initial conditions. The polynomials $O_i$ serve to encode these
initial conditions.

In more detail, we can decompose $f$ in terms of powers of $x$
modulo $n$,
\begin{equation} f_j(x) = (1-q^j)^{-1} \sum_{i\equiv
j\mod(n)} a_i x^i\end{equation}

\begin{eqnarray}
O_1(u) &= & \sum_{i<j}\tilde f(q^i x)\tilde f(q^j x)\\
&=& \sum_j \frac12\left[(f_j(1-q^j)^{-1} +\mu (1-q^{-j})R_{n-j}(u)x^j)\right.\\
&&(f_{n-j}(1-q^{-j})^{-1} +(1-q^{j})\mu R_{j}(u)x^{n-j})
\\&&\left. (q^j+q^{2j} +\dots+ q^{(n-1)j})(n-1)\right]
\end{eqnarray}
The sum appearing in the above equation can be evaluated using
$1+q^j+q^{2j} +\dots+ q^{(n-1)j}=0$. We get
\begin{eqnarray}
O_1(u) + (n-1) \frac 12 (2-q^j-q^{-j})^{-1}( f_j(x) f_{n-j}(x))
=\\\mu (n-1)\left[ \sum_j f_j(x) x^{n-j} R_j(u)-\frac { \mu u
R_j(u)R_{n-j}(u) (2-q^j-q^{-j})}{2}\right]
\end{eqnarray}
Now, the left hand side of the above equation is a polynomial in
$u$, while the right hand side is a power series in $u^{-1}$.
Equality means that the coefficients
of $u^{-k}$ of the right hand side must vanish identically.

After a little bit of algebra collecting the powers of $u^{-1}$
this gives us the following relations
\begin{equation}\label{eq:quantum1}
\sum_i a_i <\lambda^{i+ns}> = \frac
{\sum_j(2-q^j-q^{-j})<\lambda^{ns-j}><\lambda^j> }{2}
\end{equation}

Now, we will show that these equations can be derived from manipulating
loop equations in the associated matrix model. We can ask why these relations
look so different from each other for different values of $n$, and what
is special about $q$ being a root of unity in the matrix model.
This is what we will analyze in the next section. Notice that the concept
of holomorphy has allowed us to propose some polynomial relations for
twisted one point functions in the matrix model, but at this moment this
should be interpreted as a guess, since there is no proof that the 
construction of the moduli space based on matrix models necessarily gives 
the right answer.

\section{Obtaining the Quantum relations from loop equations for a
matrix model}

Consider the three matrix model whose potential is given by
\begin{equation}
W(X,Y,Z) = N\mu^{-1} [\tr( XYZ-qXZY) + \tr V(X)]
\end{equation}
with
\begin{equation}
V(x) = \sum_{i\not\equiv 0 \mod(n)}  \frac{a_i}{i} x^i
\end{equation}
a polynomial, and $q^n=1$ a primitive $n$-th root of unity.

Because the matrix model is given formally by the superpotential of
a supersymmetric field theory, we will classify the one point functions
$<f(\phi)>$ by their quantum numbers under the $\BZ_n\times \BZ_n$
quantum symmetry of the corresponding chiral operator in the quantum field
theory. A word which has charge zero under the quantum symmetry will be
called untwisted, and a word which transforms non-trivially under the quantum
symmetry will be called twisted. This nomenclature is borrowed from the fact
that the theory with $V(x)=0$ results from the field theory at a special
orbifold singularity.
 We will also assign $Y$ and $Z$ charges
$\pm 1$, so that we will also use this $U(1)_{YZ}$ charge to classify words.

Since this second charge is conserved by the potential and the measure,
 we can
consistently set to
zero the one point function
of any word whose charge is different from zero. This
constraint at the level of the
classical field theory implies that there are no fractional
branes on the singular curves of the CY, and is a simplifying assumption
to reduce the amount of algebra. We want to impose this constraint
as well on the matrix model, since we are interested in situations where the
matrix model has an isolated vacuum about which all of the modes are
massive.

We will use throughout the following simplifying notation
convention for the
one point functions of
words in the matrix model
\begin{equation}
<f(\phi)> = <\frac 1 N \tr(f(\phi))>
\end{equation}

Now we will proceed to derive the equations \ref{eq:quantum} from the
loop equations associated to this matrix model.

The universal loop equations
can be written by considering an infinitesimal change of
variables $\delta\phi^i= f( \phi) $ (one assumes this is a monomial
written as some ordering of the matrix variables, i. e. a word in the
matrices), and realizing that the
following expression is the integral of
a total derivative and therefore vanishes
\begin{equation}
 \int [d\phi]  {\partial_ {(\phi^i)}}^j_k (\delta\phi^i)_j^k \exp( -W)
=\int [d\phi]  \tr( \partial_ {\phi^i} \delta\phi^i) \exp( -W) =  0
\end{equation}
The loop equations will have the universal form
\begin{equation}
\sum \mu <f_1><f_2> =  < \frac{\partial W}{\partial\phi^i} \delta\phi^i>
\end{equation}
where the sum is over all $f_1, f_2$ monomials such that
$f_1 \phi^i f_2 = f(\phi)$, this is, from splitting the word $f$ into two words
connected by the letter $\phi^i$. The above result is using factorization
of the large $N$ limit of the matrix model.

Because we set all words with charge different than zero to zero,
we need to look only at variations of the matrices $X,Y,Z$ which have the same
charge as the object we are varying.
Thus we will look for $\delta X = f(X,Y,Z)$, where $f$ has the same number
of $Z$ and $Y$ matrices. Similarly $\delta Y = g(X,Y,Z)$ where the word
$g$ has one more field of type $Y$ than of type $Z$.

There is also the $U(1)_{R'}$ to consider. It's existence gives a 
grading to the set of words. One can systematically explore the set of 
possible monomial variations by exploiting the $U(1)_{R'}$ symmetry
and take variations with increasing $U(1)_{R'}$ charge.

\subsection{The first equation}

We will study the loop equations inspired in the solution of the two 
matrix problem given in \cite{Stau}. The main idea is to write the loop
equations in such a way that one can learn information on
 complicated words in terms of simpler words.

Consider the following matrix variations, of minimal number of $Y, Z$ pairs
\begin{equation}
\delta Y = X^m Y X^{k-m}
\end{equation}
The loop equations associated to this variation take the simple form
\begin{equation}\label{eq:cycle1}
\mu < X^m> <X^{k-m} > = < X^{m+1} Y X^{k-m}  Z - q X^m Y X^{k-m+1} Z>
\end{equation}
On the right hand side there are $k+2$ independent possible
orderings of the fields, while we only have $k+1$ different
equations. These one point functions on the right have exactly one pair of
$Y,Z$ matrices.
Variations of $Z$ of this type will produce the same equations as above,
so they do not lead to new loop equations.

Also, one can consider the variation of $X$ given by $\delta X = X^{k+1}$,
from which we get
\begin{equation}\label{eq:flip1}
\mu \sum_{m=0}^k
<X^m><X^{k-m}> - <V'(X) X^{k+1}> = <X^{k+1} YZ> -  q<X^{k+1} ZY>
\end{equation}
which gives an extra equation for some of the same
one point functions in equation
\ref{eq:cycle1}.

In total we have $k+2$ equations for the $k+2$ variables
$<X^{m} Y X^{k+1 -m}  Z>$, so in a generic case we could say that the one
point functions of $<X^{m} Y X^{k+1 -m}  Z>$ are all determined from the
one point functions of $<X^m>$. Considering the $U(1)_{R'}$ 
grading, the words involving
$Y,Z$ have higher degree than those with only $X$. 

Notice that one can use the equations to shift $Z$ to the right in the
word $<X^{m} Y X^{k+1 -m}  Z>$, and then eventually
 one comes back to the original
configuration with a phase, so in this case one can solve for any of the  one
point functions in the right hand side, as long as the phase is not
zero.

However, there are situations in which these equations are not linearly
independent. This will happen when $k$ is a multiple of $n$, and then on
transporting $Z$ around the loop the total accumulated phase is
$q^{k} = 1$. From here the equations for the vevs 
$<X^{m} Y X^{k+1 -m}  Z>$ are linearly dependent.
This means that the one point function $<X^{m} Y X^{k+1 -m}  Z>$ is not
determined in this case. Notice that this is exactly the case
when the word
is untwisted \footnote{From the point of view of the classical
SUSY gauge field theory,
this function in the chiral ring
receives contributions from the moduli space
of a probe brane, so it can take any value.}.

Consistency of the loop equations then imposes a constraint between the
elements of the left hand side. This constraint is given by summing
\begin{eqnarray}\label{eq:constraint}
\sum \mu q^{m} <X^m> <X^{k-m}> &=& \sum q^{-m}
< X^{m+1} Y X^{k-m}  Z - q X^m Y X^{k-m+1} Z>\nonumber\\
&=& q^{-k} <X^{k+1}YZ> - q < Y X^{k+1} Z>\\
&=&
<X^{k+1}YZ>-q < Y X^{k+1} Z>
\\&=&\mu \sum_{m=0}^k
<X^m><X^{k-m}> - <V'(X) X^{k+1}>\nonumber
\end{eqnarray}
and this equation can be rewritten after some manipulations
as follows
\begin{equation}
<V'(X) X^{k+1}> = \frac 12 \sum \mu ( 2-q^m -q^{-m}) <X^m><X^{k-m}>
\end{equation}
which is exactly the first of the relations that appeared in
\ref{eq:quantum1}.

As the reader can verify, the equation above is written only in terms
of twisted one-point functions, as $(2-q^m-q^{-m})=0$ whenever $m$ is
a multiple of $n$. Since $V$ is a polynomial (lets say of degree $s$),
 the above equation can be
seen as a recursion relation that determines the highest degree
component of $V'(X) X^{k+1}$, namely the one point function of $<X^{k+s}>$
(in the case where $k = r n$),
in terms of one point functions of lower degree.
The first such component that can be determined is the one
corresponding to $<x^s>$, and then all other components modulo $n$.

Now, we will solve the constraints for the one point functions
$<X^{k+1}YZ>$ for $k$ not a multiple of $n$,
 because we will need them for future calculations.
This is easily  seen to be given by  solving \ref{eq:constraint} when
$k$ is not a multiple of $n$. In this case there is no equality between the
second and third lines, and instead we get two linearly independent
equations with two unknowns.
We get that
\begin{eqnarray}\label{eq:twistedsub}
(q^{-k}-1)<X^{k+1}YZ> & = & \sum \mu q^{m} <X^m> <X^{k-m}>
\nonumber \\&&-\mu \sum_{m=0}^k
<X^m><X^{k-m}> + <V'(X) X^{k+1}>\\
&=& \mu\sum (q^m-1)<X^m><X^{k-m}> + <V'(X) X^{k+1}>
\end{eqnarray}
Notice that on the right hand side of the third line we do get
untwisted one point functions, so these can in principle 
receive contributions
from branes in the bulk. This happens when
$k-m$ is a multiple of $n$, or from $V'(X)X^{k+1}$ when certain terms in
the potential are non-zero.

\subsection{The second Quantum equation}

The next level of difficulty is obtained
by taking words of next to minimal numbers of pairs of $Y,Z$ fields.
For example
$\delta Y = Y X^m Z X^n Y$, and all possible orderings of such
monomials. Although it looks much harder to solve this linear system,
we do not need to study the most general variations of $Y$ of this type.
The main effect of the loop equations is to tell us what happens when we
permute two of the symbols in a word in terms of one point functions of lower
numbers of pairs of $Y,Z$. Thus any one point function can be determined
from any other permutation by a series of steps. We only need to worry
about the possibility of the system of equations not being
linearly independent, just as we saw previously.
This can happen when the words we study are
untwisted, and then one produces a constraint between words of smaller
degree.

To check this redundancy, we just need to take an initial word
$<X^{k+2} YYZZ>$ and cycle the $Z$ around the $X$ and $Y$.
We will return with the same phase to the initial configurations when
$k$ is a multiple of $n$. This is because passing a
$Z$ through an $X$ gives a factor of
$q$, and passing a $Z$ past a $Y$ gives a factor of $q^{-1}$, so the
total phase from cycling $Z$ around the trace is $q^{k+2} q^{-2} = q^k$.

We only need to look at the variations $\delta Y = X^{m} Y^2 Z
X^{k+1-m}$, and their corresponding loop equations
\begin{equation}\label{eq:cycle2}
\mu <X^m><YZ X^{k-m}>
= <X^{m+1}Y^2ZX^{k+1-m}Z> - q < X^m Y^2 Z X^{k+2-m} Z>
\end{equation}
where we have set the charged one point functions to zero.

We also need two equations to cycle the $Z$ past the $Y$. These are
provided by the variations $\delta X = YZ X^{k+2} $ and
$\delta X = Z X^{k+2} Y$. These two loop equations will look
as follows \footnote{We write only the equations with $<X^kZ>=0$
for all $k$}
\begin{eqnarray}\label{eq:flip2}
 <YZ YZ X^{k+2}> - q <ZYYZX^{k+2}> &= &\nonumber \\\mu \sum_m
<YZ X^m><X^{k+1-m}>
- <YZ X^{k+2} V'(X)> \nonumber &&\\
 <YY ZZ X^{k+2}> - q<YZYZ X^{k+2}> &=&\nonumber\\
-<Z X^{k+2} Y V'(X)>
\end{eqnarray}

Summing \ref{eq:cycle2}  we get, in the case where
$k$ is a multiple of $n$ that
\begin{eqnarray}
&<X^{k+2}Y^2ZX^{k+1-m}Z> - q^{2} <  Y^2 Z X^{k+2} Z> =&\nonumber \\
&\mu \sum_{m=0}^{k+1} q^{k+1-m} <X^m><YZ X^{k-m}>
\end{eqnarray}
which is also the sum of the equations in \ref{eq:flip2}
\begin{eqnarray}
&<X^{k+2}Y^2ZX^{k+1-m}Z> - q^{2} <  Y^2 Z X^{k+2} Z> =&\nonumber\\
&q\mu \sum_m <YZ X^{k+1-m} ><X^{m}>&\nonumber \\
&-q <YZ X^{k+2} V'(X)> -<Z X^{k+2} Y V'(X)>&
\end{eqnarray}
We can rewrite the constraint as follows
\begin{eqnarray}\label{eq:preQuantum2}
&q \mu \sum_m (1-q^{-m}) <X^m><YZ X^{k+1-m}>= &\nonumber\\
&q <YZ X^{k+2} V'(X)> +<Z X^{k+2} Y V'(X)>&
\end{eqnarray}
Again, we notice that the one point functions of $<X^m>$ for $m$ a
multiple of $n$ do not contribute, because $q^m=1$. This means
the terms proportional to $\mu$
only depends on twisted one point functions.
Notice that the right hand
side also only depends on
twisted one point functions, at least as far as the powers
of $X$ that are excluded from $V(X)$.
Notice that the constraint above is not in the form which we want,
which depends
only on one point functions of the type $<X^m>$. However, we can use
the results from the previous subsection to express all of the one
point functions appearing in the above equations in terms of one point
functions of $<X^m>$, possibly including the ones where $m$ is a multiple of
$n$, which we have no way of determining from our previous arguments.
The equations obtained in this manner will contain
terms of order $\mu^2$ (triple trace), of order $\mu$ (double trace), and
of order $\mu^0$ (single trace). We still have to check that after
we do all of these substitutions,  the
one point functions appearing in the above equation are all twisted, and that
the equation coincides with what we have shown already.
This would lend support to the existence of a recursion relation
determining only the twisted powers of $X$.

At this point we need to substitute explicit forms for $V'(X)$
in terms of it's polynomial expansion, because the way we will substitute in
equation \ref{eq:twistedsub} depends on what particular
powers of $X$ are available. We also need to do an extra step in
equation \ref{eq:preQuantum2} and move the $Z$ past some of the $X$
in the one point function of $<Z X^{k+2} Y V'(X)>$.

Consider expanding $V'(X)=
\sum_{i\not\equiv 0 \mod(n)} a_i X^{i-1} $. It is clear that we will get
one point functions of monomials of the type
$<Y X^{i-1} Z X^{k+2}>$, in the case where $k$ is a multiple
of $n$. The following equality is easy to show for these one point functions
by summing terms of the equations \ref{eq:cycle1}.
\begin{equation}
<YZ X^{i+k+1}>- q^{i-1}  <Y X^{i-1} Z X^{k+2}>=
\mu \sum_{m=0}^{i-2} q^m <X^m><X^{k+i+1-m}>
\end{equation}
This can be used to rewrite
\begin{equation}
<Y X^{i-1} Z X^{k+2}> = q^{1-i}
\left[<YZ X^{i+k+1}>-
\mu \sum_{m=0}^{i-2} q^m <X^m><X^{k+i+1-m}>\right]
\end{equation}
From here we can check that the coefficient of $a_i<YZX^{k+1+i}>$
in equations \ref{eq:preQuantum2} coming from the potential is given by
$q+q^{i-1}$.

We will now only check that the terms
of order $\mu^0$,  which are untwisted and
proportional to the second power of the coefficients in the potential
$(a)^2$ are canceled. For this, we need to use equation
\ref{eq:twistedsub} and only keep terms that involve two powers of the
$a$.

Indeed, we get the sum
\begin{eqnarray}
\sum_i (q+q^{1-i}) a_i <YZ X^{k+i+1}>)
&=& \sum_i (q+q^{1-i})(q^{-(k+i)}-1)^{-1}
a_i <X^{k+i+1}V'(X)> +O(\mu) \nonumber \\
&=&\sum_{i,j} (q+q^{1-i})(q^{k+i}-1) a_i a_j <X^{k+i+j}>+O(\mu)
\end{eqnarray}
Now we want to consider the case where $i+j$ is a multiple of $n$, and we
are summing two different terms with the coefficient $a_i a_j<X^{k+i+j}>$.
Also, remember that $k$ is a multiple of $n$, so we can drop it from
the powers of $q$.
The coefficient for the sum of these two terms is equal to
\begin{equation}
q(1+q^{-i})(q^{-i}-1)^{-1} +q(1+q^{-j}) (q^{-j}-1)^{-1}
\end{equation}
where we substitute $i= -j\mod n$, so that the numerator of the
expression is equal to
\begin{equation}
q( 1+q^i)(q^{-i}-1)+q(1+q^{-i})(q^i-1)
=q(q^{-i}-q^i + q^i-q^{-i})=0
\end{equation}

This shows that there are no untwisted powers of $X$ appearing in the one
point functions of this second constraint which are of second order in the
$a_i$. As these are also the terms that contain
the highest power of $X$ that is available. This shows that
the above
equations will determine higher twisted one-point functions of $X$.

We still have not shown that in the double trace and triple trace operators
one can eliminate all the untwisted powers of $X$.
Since we have already found relations using the
matrix theory calculation of the moduli space, and moreover we have
shown that the constraints give the right result
for the first quantum equation plus a consistency condition for the
second equation, we will leave
this lengthy  algebraic  calculation as an exercise
to the reader.

The methodology for determining the recursion
relations should now be evident to the reader. One has to add
more and more pairs of $Y,Z$ operators, and consider variations of the form
\begin{equation}
\delta Y \sim X^{k+s-m} Y^{s+1}Z^s X^m
\end{equation}
which allow us to circulate one $Z$ past all the $X$. We also need to consider
the variations
of $X$ that allow us to flip $Z$ past the $Y$ in words with charge $0$.
Again, when the words are untwisted, the words will be undetermined
classically and
the loop equations are linearly dependent. Consistency imposes constraints
that will involve twisted words with smaller
numbers of pairs of $Y,Z$. Since these words are twisted, they will belong
to the set of relations that can be solved in terms of smaller numbers of
pairs of $Y,Z$ variables. One should be able to show that after performing
some amount of algebra one can reduce the system so that there are only
one point functions for powers of $X$ which are untwisted. The calculations
at each succeeding step become more involved
\footnote{It would be interesting to find a more effective
method to calculate these relations in the matrix model to give a more
compelling proof, say in terms of eigenvalue distributions for $X$.
 The author has not found a better way to understand it however.},
 but this is a
recursive algorithm to get the constraints.

Summarizing, we see that the presence of a moduli space of vacua has 
given us a degenerate system of linear equations for the one point 
functions $<X^K Y^M Z^{M-1}X^{K'}Z>$. For generic values of $q$ the 
associated linear equations determine these one point functions
completely in terms of functions of lower degree. However, one does not get 
constraints that involve one point functions of $X$ alone. These should be 
considered as unknown parameters determined by the ``infinite number of cuts''
 of the matrix model.
For the special 
values of $q$ that we studied, this does not happen, and the linear 
system is degenerate. 
For this linear system to have a solution, the terms of lower degree
must satisfy a constraint. It is exactly these 
constraints that encode the shape of quantum moduli space of 
vacua of the associated SUSY gauge theory. Moreover, the presence of only a 
finite number of singularities in moduli space guarantee that for certain 
generalized resolvents, one only has a finite number of cuts.
Also, due to the special 
form of the potential, in the particular case we are studying, the 
solution effectively reduces to a one matrix model whose solution is 
determined by a curve 
$Q_{quatum}(t,u)$.

\section{Partial gaugino condensates}

We have now described a partial solution of the three matrix problem, and we
have found that some interesting generalized resolvents
(the $R_j(u)$) can be obtained from the curve $\tilde Q(t,u)$.

In particular, it is interesting to ask how many eigenvalues in the
matrix model are located around each classical saddle point of the
theory. In the gauge theory this will be
 the calculation that determines the
gaugino condensates from the geometry.

Now we want to relate the $R_j(u)$ to
the problem of the distribution of eigenvalues of $X$.
In particular, $R_j(u)$ only has information about
the distribution of fractional branes, so it only contains information about
the eigenvalues of $X$ associated to the fractional branes.

Let us look at the definition of $R_{n-1}(u)$. It is given by
\begin{equation}
R_j (u) =  \frac 1 N \sum_i\frac{\lambda_i^{j}}{u-\lambda_i^n}
= \sum_{k=0}^\infty\frac {< \lambda^{nk+j}>}{u^{k+1}}
\end{equation}
Classically, this function has poles at the eigenvalues
$\lambda_i^n$. It is more convenient
to introduce a cover of the
$u$ plane where $u=\gamma^n$, and then write
\begin{equation}
R_j(\gamma^n) = \frac 1 N \sum_i\frac{\lambda_i^{j}}{\gamma^n-\lambda_i^n}
\end{equation}
In particular, if we choose $j=n-1$, the residue of the poles associated to
a particular eigenvalue $\gamma= \lambda_i$ is $1/nN$.

However, we have to take into account that classical saddle points for the
eigenvalues associated to fractional branes $R^+_\alpha$ come located also at
$qx,q^2x,\dots q^{r-1}x$ where $r$ is the dimension of the associated
representation. And there is a conjugate representation $R^-{\alpha}$ that
appears as well for the other set of eigenvalues $q^rx, \dots q^{n-1}x$.

So if we want to count how many eigenvalues $N_\alpha$ in the matrix model
are located around a
classical saddle point which has an eigenvalue located at $x_0$,
we want to do the following contour
integral
\begin{equation}
\frac{1}{2\pi i}\oint_C R_{n-1}(\gamma^n)d\gamma
= \frac{N^+_\alpha-N^-_\alpha}{nN} (1+q^{-1}+\dots q^{-r+1})
\end{equation}
where $C$ is a contour over the eigenvalue plane which contains
$x_0$.
Notice that the above expression only depends on $N^+_\alpha-N^-_\alpha$
which is the total number of fractional branes that can not be removed away
from the singularity classically.
 If one studies the Klebanov Strassler system, this
would count
exactly the difference between the ranks of the gauge groups, which is an
invariant under the duality cascade.

Giving specific values to $N_{\alpha}/N$ should determine the curve in the
matrix
model, by asking which curve has the right residues for a given value of
$\mu$.

In the gauge theory, the generalized resolvent $R_{n-1}$ in the
matrix model is related to the generalized resolvent for the chiral ring
given by
\begin{equation}
{\cal R}_j(\gamma^n)
= \tr( \frac{X^j W^2}{\gamma^n-X^n})
\end{equation}

The gauge invariant definition of the partial gaugino condensate
$S_i$ associated to the fractional branes at some conifold singularity
should be
given by the following expression
\begin{equation}
S_\alpha
 = n\frac{1}{2\pi i(1+q^{-1}+\dots q^{-r+1}}\oint_C {\cal R}_{n-1}
(\gamma^n)d\gamma
\end{equation}
where the contour is around one of the eigenvalues of $X$ at the given
singularity, and $S_\alpha$ is the partial gaugino condensate
of the full collection of fractional branes at the singularity,
once we get to the bottom of the duality cascade of the local conifold.
This expression should be compared with the recent setup of Dijkgraaf
and Vafa based on supergroups \cite{DVd}, where everything only depends on the
``difference of the gaugino condensates''. That statement is equivalent
to branes in the bulk not contributing to the deformations of the geometry, as
seen also in \cite{Bq}.

From here, again, it should follow that one should be able to determine the
deformation of the geometry caused by the fractional branes.

It would also be interesting to determine whether the above computation 
coincides
with calculating periods of the holomorphic three-form of the associated CY
geometry. This problem will be left for future study.

\section{Conclusion}

We have seen in this paper that quantum deformed moduli spaces for
geometrically engineered theories encode very non-trivial
relations in the associated matrix models. The relations can be
argued to be  consistency conditions for the loop equations of the
matrix model, when the loop equations degenerate.
Indeed, the constraints are powerful enough to give a curve
which solves the matrix model for a generic number of cuts.

It is not clear under what circumstances the approach presented in
this paper is valid. The most conservative possibility is that it
applies only to those theories that correspond to a geometric
realization of branes at singularities. This is suggested because
the effects of the probe brane on fractional branes (what one
would call the back reaction) cancels. It is very likely that
this is tied to the fact that the low energy effective field
theory of the probes on the moduli space is essentially ${\cal
N}=4$ SYM, and therefore it does not get corrected, nor does it
develop a gaugino condensate. To this extent one can argue that
their treatment should be essentially classical. This does not
prevent the fractional branes from affecting the geometry of the probe,
 and we
see that indeed they do by forcing geometric transitions. These
transitions exactly deform the points in moduli space where the
probe brane effective field theory has less supersymmetry.

The method serves to solve the theory for values of $q$ which are
roots of unity. Since in the SUSY gauge theory $q$ is in principle
a complex variable, one can expect that the solution of vevs of
chiral operators depends holomorphically on $q$, with possible
singularities at some special values of $q$. These singularities
should happen exactly when the associated $SU(2)_q$
representations constructed in the paper 
disappear (at roots of unity). For a given rank of
the gauge theory $N$, there are only classically a finite number of values of $q$
which can be pathological (those that are $r$-roots of unity with
$r\leq N$, or $q=0,\infty$). In principle this information should 
determine the pole structure
of the holomorphic observables of the gauge theory as a function
of $q$. Since we can in principle solve the theory for roots of
unity, one should be able to use analyticity (holomorphy of the
supersymmetric gauge theory) to solve the theory for any $q$, and
address the issues of multi-cut solutions of the ${\cal N}=1^*$
gauge theory. 
These and other related issues are currently under
investigation.

\section*{Acknowledgements}

I would like to thank F. Chacazo, M. R. Douglas, D. Gross, A.
Hashimoto, V. Kazakov, N. Seiberg, M. Staudacher, E. Witten for
various discussions related to this subject. I would also like to
thank  the Rutgers theory group for their hospitality while this
work was developed. Research supported in part by DOE grant
DE-FG02-90ER40542.

\section*{Appendix}

In this appendix we treat the three matrix model of the paper by
considering a saddle point for the eigenvalues of the matrix $X$.

The potential of the theory is given by $$N\mu^{-1}(
\tr(XYZ-qXZY)+\tr(V(X)))$$
  After one diagonalizes $X$, with eigenvalues $\lambda_i$
  and if one integrates $Y,Z$, one obtains
an effective potential for the eigenvalues of $X$ given by
\begin{equation}
N\mu^{-1} (\sum_i V(\lambda_i)) +\sum_{i<j}[
-2\log(\lambda_i-\lambda_j)+\log(\lambda_i-q\lambda_j)
+\log(\lambda_i-q^{-1}\lambda_j)]
\end{equation}

The saddle point equations for the eigenvalues $\lambda$ are given
by
\begin{equation}
V'(\lambda_i) +2 \mu N^{-1} \sum_{j\neq i}
[-\frac{2}{\lambda_i-\lambda_j}+\frac1{\lambda_i-q\lambda_j}
+\frac1{\lambda_i-q^{-1}\lambda_j}]
\end{equation}

This is effectively a one matrix model for $X$. Notice that in the
limit where the eigenvalue distribution becomes continuous, for
each cut in the eigenvalue plane associated to $X$, there are
image cuts in the complex plane rotated by $q$ and $q^{-1}$ in the
saddle point equation for the eigenvalues. For the $SU(2)_1$
representations one has eigenvalues on these images, and these
give a picture of the $SU(2)_q$ representations. In the case of
$q$ a root of unity, there is a bound on the number of images per
cut, and this makes the case of a root of unity special.

Also, one can derive the first quantum equation \ref{eq:quantum1} for the
case $q^n=1$ by multiplying the above saddle point equation by
\begin{equation}
\frac {\lambda_i}{u-\lambda_i^n}
\end{equation}
summing over all the eigenvalues, and dividing by $N$.
However, the author has found no systematic way to extend this procedure to
derive the other quantum equations 
from the above saddle point equations.

\end{document}